\documentclass[10pt,journal,final,twocolumn]{IEEEtran}
\usepackage{graphics}
\usepackage{epsfig}
\usepackage{balance}
\usepackage{tabularx}
\usepackage{multirow}
\usepackage{subfigure} 
\usepackage{color,xcolor}
\usepackage{cite}

\usepackage{amsmath,amssymb,amsfonts}
 \usepackage{amsmath,amsfonts,amsthm,bm}
\usepackage{algorithm}
\usepackage{algpseudocode}

\begin{document}

%

%
\title{{\color{black}  One-Bit ADCs/DACs based MIMO Radar: Performance Analysis and Joint Design}  }
%
%
%

\author{Minglong~Deng,~Ziyang~Cheng, \IEEEmembership{Member, IEEE},~Linlong~Wu, \IEEEmembership{Member, IEEE},\\~Bhavani~Shankar, \IEEEmembership{Senior Member, IEEE}, and~Zishu~He \vspace{-2em}
\thanks{This work is supported by XXX. \emph{Corresponding author: Ziyang Cheng}. }
\thanks{Minglong Deng, Ziyang Cheng and Zishu He are with the School of Information and  Communication Engineering, University of Electronic Science and Technology of China (UESTC), Chengdu 611731, China. E-mail: zycheng@uestc.edu.cn. }

\thanks{Linlong Wu and Bhavani Shankar are with the Interdisciplinary Centre for Security, Reliability and Trust (SnT), University of Luxembourg, Luxembourg City L-1855, Luxembourg. E-mail: \{linlong.wu, bhavani.shankar\}@uni.lu. Their work is supported in part by ERC AGNOSTIC under grant EC/H2020/ERC2016ADG/742648 and in part by FNR CORE SPRINGER under grant C18/IS/12734677.}
}
\maketitle

\begin{abstract}
	\textcolor{black}{Extremely low-resolution (e.g. one-bit) analog-to-digital converters (ADCs) and digital-to-analog converters (DACs) can substantially reduce hardware cost and power consumption for MIMO radar especially with large scale antennas. 
In this paper, we focus on the detection performance analysis and joint design for the MIMO radar with one-bit ADCs and DACs. 
Specifically, under the assumption of low signal-to-noise ratio (SNR) and interference-to-noise ratio (INR), we derive the expressions of probability of detection ($\mathcal{P}_d$) and probability of false alarm ($\mathcal{P}_f$) for one-bit MIMO radar and also the theoretical performance gap to infinite-bit MIMO radars for the noise-only case.  We further find that for a fixed $\mathcal{P}_f$, $\mathcal{P}_d$ depends on the defined quantized signal-to-interference-plus-noise ratio (QSINR), which is a function of the transmit waveform and receive filter. 
Thus, an optimization problem arises naturally to maximize the QSINR by joint designing the waveform and filter. For the formulated problem, we propose an alternatin\emph{g} wavefo\emph{r}m and filt\emph{e}r d\emph{e}sign for QSINR maximiza\emph{t}ion (GREET). At each iteration of GREET, the receive filter is upadted via the minimum variance distortionless response (MVDR) method, and the one-bit waveform is optimized based on the alternating direction method of multipliers (ADMM) algorithm where the closed-form solutions are obtained for both the primary and slack variables. Numerical simulations are consistent to the theoretical performance analysis and demonstrate the effectiveness of the proposed design algorithm.
}
\end{abstract}


%
\IEEEpeerreviewmaketitle

\section{Introduction}
%
%
%
%

\IEEEPARstart{D}{uring} the past few years,  {\color{black} multiple-input multiple-output} (MIMO) radar, as an enhanced version of traditional phased-array radar, has received great attention due to its more flexible transmit scheme\cite{A1,A2,A3,A4,A5,A6,A7,A8,A9}. 
Compared with the phased-array radar which emits   a single waveform, the  MIMO radar  transmits  multiple waveforms that may be correlated or uncorrelated with each other. 
Such waveform diversity provides  MIMO radar with many performance improvements, such as  enhanced parameter identifiability \cite{A2}, higher direction-of-arrival (DOA) estimation accuracy \cite{A3,A4}, better transmitting behaviors \cite{A5,A6,A7}, as well as improved ability of interference suppressing \cite{A8,A9}.

Waveform design is one of the most effective ways to achieve above mentioned performance improvements. 
Many performance metrics have been taken into account  to design the MIMO radar  waveforms, such as the {\color{black} Cram\'{e}r-Rao} bound (CRB) \cite{B1,B2}, mutual information (MI) \cite{B3,B4,B6},  spectral shaping \cite{rowe2014spectrally,jing2018spectrally,wu2019sequence},  beampattern synthesis  \cite{A5,A6,A7,B7,B8,B9}, and active sensing \cite{6104176,6472022}, to name just a few.
Particularly, to improve the performance of target detection and parameter
estimation for MIMO radar, the maximization of output signal-to-interference-plus-noise ratio (SINR)  is often raised as a figure of merit  to design the waveform \cite{A8,A9,B10,B11,B12,wu2017transmit}.
For example, in the presence of signal-independent interference, the maximization of detection probability is considered to design the optimum code matrix in \cite{A8}. 
In the presence of signal-dependent interference, \cite{A9} proposes a gradient based algorithm to jointly design the transmit waveform and receive filter by maximizing the SINR with a priori knowledge of the target and clutter statistics. 
In addition, \cite{B11} proposes two sequential optimization algorithms (SOA) based on the semidefinite relaxation (SDR) method to the maximization of SINR while keeping the constant modulus (CM) constraint as well as a similarity constraint. 
In \cite{B12}, the CM constraint is relaxed to a general peak-to-average-power ratio (PAPR) constraint, and a block successive upper-bound minimization method of multipliers framework \cite{BS1} is devised to solve the resultant problem. 
The work \cite{wu2017transmit} also considers the joint design problem and derives an algorithmic framework, under which all the above mentioned constraints can be readily handled.

Although the theoretical research of MIMO radar has been studied very thoroughly, the practical implementation of MIMO radar still faces great challenges. For instance, all the above-mentioned methods design the waveform and receive filter by implicitly assuming ideal analog-to-digital
converters (ADCs) and digital-to-analog converters (DACs). 
{\color{black} However, using high-resolution ADCs/DACs will considerably increase the circuit power consumption as well as the hardware cost, especially for MIMO radar with large bandwidth, high sampling rate, and large scale antenna array \cite{761034}.}
To overcome such difficulties, researchers attempt to employ the sub-Nyquist technology, which samples the signal at a rate less than the Nyquist rate \cite{C1,C2,C5,C6}.
For instance, in \cite{C1}, based on the compressive sampling (CS) technology \cite{C7}, the angle and Doppler information on potential targets from temporally and spatially under-sampled observations are obtained.
In \cite{C2}, a sparse array design framework is proposed for MIMO radar allowing for a remarkable reduction in the number of antennas, while still attaining performance comparable to that of {\color{black} a fully filled (Nyquist) array}.

In contrast to the sub-Nyquist technology, using low-resolution ADCs/DACs is a more efficient way and is attracting considerable critical attention, because the circuit power consumption decreases exponentially with decrease in the number of quantization bits \cite{761034}. Particularly, recent developments in the field of communication and radar have led to a significant interest in the utilization of one-bit ADCs/DACs \cite{D2,D3,D4,D5,D8,D9,E1,D6,E2,E3,9406387,9585542,E4,E5}.
Specifically, for a radar system with one-bit ADCs/DACs, the studies involved include  target imaging \cite{D8}, target detection \cite{D9,E1}, parameter estimation \cite{D6,E2,E3,9406387,9585542}, and waveform design \cite{E4,E5}.
For example, the one-bit likelihood ratio test (LRT) detector is derived and the detection performance is evaluated in \cite{E1}. 
To get the angle and velocity of a target, \cite{E2} proposes a maximum likelihood (ML) method to recover the virtual array data matrix from one-bit observations, and then estimates the parameters by using the resulting data.
More recently, the performance of DOA estimation for a sparse linear array with one-bit ADCs is investigated in \cite{9585542}.
On the other hand, when employing one-bit DACs, it leads to a problem analogous to the nonlinear precoding in communication \cite{D2,D3} and binary waveform design in radar \cite{E4,E5}.
For example, in \cite{E4}, a fast Fourier transform (FFT) based framework is proposed to optimize the one-bit transmit waveform by minimizing the MSE between the designed and desired beampatterns. 
In \cite{E5}, the one-bit transmit sequences are designed to ensure the DOA estimation accuracy of a dual-function radar-communication (DFRC) system equipped with one-bit DACs.  

Note that, a number of critical questions remain about the development of one-bit MIMO radar.
For example, although there are many studies in the literature on the interference suppression by jointly designing the transmit waveform and receive filter, such as \cite{A9,B11,B12,wu2017transmit}, most are restricted to MIMO radar with infinite-bit ADCs/DACs. 
In the context of one-bit MIMO radar, what remains unclear is how one-bit ADCs/DACs affect the detection performance and joint design of MIMO radar in the presence of interference, which motivates us to carry out this work. 
Specifically, the main contributions and novelties of this paper are summarized as follows:

\begin{itemize}
	\item Based on the low input signal-to-noise ratio (SNR) and interference-to-noise ratio (INR) assumption, the detection performance of one-bit MIMO radar, including the probabilities of detection and false-alarm, is derived as a function w.r.t. the transmit waveform and receive filter.
	We refer to this function as quantized SINR (QSINR), and show that the optimum detection performance in the low input SNR/INR regime can be achieved via maximizing the QSINR. 
	Then, the problem of joint design is formulated by maximizing the QSINR subject to a binary constraint. 
	
	\item In noise-only scenario, compared to traditional MIMO radar with $\infty$-bit ADCs and DACs, the possible performance loss caused by using one-bit ADCs and DACs is theoretically analyzed from the QSINR point of view.

	\item  In order to tackle the resultant problem efficiently, we propose an alternatin\emph{g} wavefo\emph{r}m and filt\emph{e}r d\emph{e}sign for QSINR maximiza\emph{t}ion (GREET). At each iteration of GREET, the receive filter is updated via the minimum variance distortionless response (MVDR) method \cite{add3}, and the one-bit waveform is optimized based on the alternating direction method of multipliers (ADMM) algorithm \cite{BS1} where the closed-form solutions are obtained for both the primary and slack variables.
	
	
\end{itemize}
Finally, several simulation experiments are presented to illustrate the detection performance of one-bit MIMO radar as well as the effectiveness of GREET in terms of the output QSINR, computational efficiency, and convergence performance.

The rest of this paper is organized as follows. 
In Section II, we give the signal model. 
In Section III, the detection performance of one-bit MIMO radar is evaluated.
In Section IV, we give the definition of QSINR.
In Section V, we introduce the optimization method named GREET.
Finally, simulation results and conclusions are given in Section VI and VII, respectively.

\textit{Notation:} Throughout this paper, the italic letter $a$ (or $A$), boldface letter $\mathbf{a}$, and upper case boldface letter $\mathbf{A}$ are denoted as a scalar, a vector, and a matrix, respectively.
The  conjugate, transpose, and conjugate transpose operators are given by $(\cdot)^*$, $(\cdot)^T$, and $(\cdot)^H$, respectively.
The imaginary unit is given by $\jmath = \sqrt{-1}$, and $||\cdot||_2$ denotes the $\ell_2$ norm of a vector.
Script letters $\mathcal{U}$,  $\mathcal{N}$, and $\mathcal{CN}$  are used for denoting uniform,  Gaussian, and {\color{black} circular} complex Gaussian distributions.
The operators $\text{vec}\{\cdot\}$, $\odot$, and $\otimes$ represent the vectorization operation, element product, and Kronecker product.
The statistical expectation and variance operators are given by $\mathbb{E}\{\cdot\}$ and $\mathbb{D}\{\cdot\}$ , respectively.
The $\Re\{ \cdot \}$ and  $\Im\{ \cdot \}$ are employed to get the real and imaginary parts of a complex variable. 
The sets of real and complex numbers are denoted by $\mathbb{R}$ and $\mathbb{C}$.
{\color{black} The italic letter $e$ is Euler's number and the function ${\rm exp}(x)$ denotes the exponential function $e^x$. Finally, we use $\mathbf{1}_N$ ( $\mathbf{0}_N$)  to denote an $N\times 1$ vector with all elements being $1$ ($0$), and use $\mathbf{I}_N$ to denote an $N\times N$ identity matrix. }

{\color{black}  \section{Signal Model}}

We consider a collocated  MIMO radar system equipped with $N_t$ transmit and $N_r$ receive antennas. Suppose that there exists one point target and $ K $ signal-dependent interferences (e.g. clutters), the baseband receive signal $\mathbf{X}\in \mathbb{C}^{N_r \times L}$  at the receive array can be expressed as \cite{B11}
\begin{equation}
 \begin{aligned}
        \mathbf{X} = \xi_0 \mathbf{a}_r(\theta_0) \mathbf{a}_t^T(\theta_0) \mathbf{S} 
        + \sum\limits_{k = 1}^K \xi_k \mathbf{a}_r(\theta_k) \mathbf{a}_t^T(\theta_k) \mathbf{S} + \mathbf{V},
\end{aligned}
\end{equation}
where 
\begin{itemize}
    \item $\mathbf{S}\in \mathbb{C}^{N_t \times L}$ denotes the discrete transmit waveform matrix, $\mathbf{V}\in \mathbb{C}^{N_r \times L}$ is the noise signal, and $L$ denotes the number of samples within a pulse width;
    \item $\theta_0$ and $\xi_0$ denote the direction and complex reflection coefficient of the target of interest, and $\theta_k$ and $\xi_k$ ($k=1,2,...,K$) denote the direction and complex amplitude of the $k$-th interference;
    \item  $\mathbf{a}_r(\theta) \in \mathbf{C}^{N_r \times 1}$ and $\mathbf{a}_t(\theta) \in \mathbf{C}^{N_t \times 1}$ are the receive and transmit steering vectors, which can be given by
\begin{subequations}
	\begin{equation}
	\mathbf{a}_r(\theta)=\frac{1}{\sqrt{N_r}}  {\color{black} [e^{-\jmath2\pi d_{r,1} \sin \theta/\lambda},...,e^{-\jmath2\pi d_{r,N_r} \sin \theta/\lambda}]^T, }
	\end{equation}
	\begin{equation}
	\mathbf{a}_t(\theta)=\frac{1}{\sqrt{N_t}}    {\color{black} [e^{-\jmath2\pi d_{t,1} \sin \theta/\lambda},...,e^{-\jmath2\pi d_{t,N_t} \sin \theta/\lambda}]^T, }
	\end{equation}
\end{subequations}
   \item  $\lambda$, $\{d_{r,1}, d_{r,2},...,d_{r,N_r}\}$, and $\{d_{t,1}, d_{t,2},...,d_{t,N_r}\}$ denote the signal wavelength, the reference positions of receive antennas, and the reference positions of transmit antennas, respectively.
\end{itemize}

Taking vectorization of $\mathbf{X}$  yields
\begin{equation}
	{\bf{x}} = {\rm vec} \{ \mathbf{X} \} = {\xi _0}{\bf{A}}\left( {{\theta _0}} \right){\bf{s}} + \sum\limits_{k = 1}^K {{\xi _k}{\bf{A}}\left( {{\theta _k}} \right){\bf{s}}}  + {\bf{v}},
\end{equation}
where  ${\bf{s}} = {\rm vec} \{ \mathbf{S} \}$,
${\bf{A}}\left( \theta  \right) = {{\bf{I}}_L} \otimes {{\bf{a}}_r}\left( \theta  \right){\bf{a}}_t^T\left( \theta  \right)$, and ${\bf{v}} =  {\rm vec} \{ \mathbf{V} \}$.
Besides, the noise ${\bf{v}}$ is assumed to obey  the complex Gaussian distribution, i.e., $\mathbf{v} \sim \mathcal{CN}(\mathbf{0}_{N_rL},\sigma^2\mathbf{I}_{N_rL})$.

In order to reduce hardware cost remarkably, in our considered model, we assume  that the transmit and receive arrays are equipped with one-bit DACs and one-bit ADCs. Specifically, each receive antenna is configured with two one-bit quantizers  to quantize the real and imaginary parts of a complex signal.  
For the transmit end, with the assumption of unit energy, the transmit alphabet  $ \cal X $ is $ \frac{1}{{\sqrt {2{N_t}L} }}\{ 1 + \jmath,1 - \jmath, - 1 + \jmath, - 1 - \jmath\} $ \cite{E4,E5}. 
While for the receive array, the quantized output of the baseband receive signal is expressed as\footnote{\color{black}Herein, we assume that the radar receiver  and the transmitter
are   sampling synchronous, and that the rate of the ADC  is equivalent to that of the DAC, such that the   number of measurements at the receiver equals to the length of the transmit waveform.}
\begin{equation}
	{\bf{y}} = {\cal Q}\left( {\bf{x}} \right),
\end{equation}
{\color{black} where ${\cal Q}\left( \cdot \right) \triangleq	{\rm sign}(\Re\{ \cdot \}) + \jmath 	\rm{sign}(\Im\{ \cdot \} )  $ denotes the complex-valued one-bit quantization function, and ${\rm sign(\cdot)}$ is the signum function. }

\color{black}
Defining the linear receive filter ${\bf{w}} \in {\mathbb{C}^{{N_r}L \times 1}}$, the output signal can be given by $z=\mathbf{w}^H\mathbf{y}$.
Generally, there has two possible states of the target, i.e., $\mathcal{H}_1$: the target is present, and $\mathcal{H}_0$: the target is absent.
The corresponding output signal under the two hypotheses is denoted as
\begin{equation}
	z = \begin{cases}
	 \mathbf{w}^H \mathcal{Q}( \mathbf{h}_1 + \mathbf{v}),& \text{under } \mathcal{H}_1, \\
	 \mathbf{w}^H \mathcal{Q}( \mathbf{h}_0 + \mathbf{v}),&\text{under } \mathcal{H}_0,
	\end{cases}
\end{equation}
where  ${{\bf{h}}_1} = {\xi _0}{\bf{A}}\left( {{\theta _0}} \right){\bf{s}} + \sum\nolimits_{k = 1}^K {{\xi _k}{\bf{A}}\left( {{\theta _k}} \right){\bf{s}}} $ and ${{\bf{h}}_0} = \sum\nolimits_{k = 1}^K {{\xi _k}{\bf{A}}\left( {{\theta _k}} \right){\bf{s}}} $.

For target detection, since $z$ is a complex number, the modulus or the square modulus of $z$ should be compared with a detection threshold $\mathcal{T}$. Here, using $\tilde{z} = |z|$ as the detection statistic yields a binary hypothesis test
\begin{equation}
\label{eq6}
	\tilde{z}  \underset{\mathcal{H}_0}{\overset{\mathcal{H}_1}{\gtrless} } \mathcal{T}.
\end{equation}
Since the statistical characteristic of $\tilde{z}$ depends on the transmit waveform $\mathbf{s}$ and receive filter $\mathbf{w}$, it is obvious that the detection performance is also associated with $\mathbf{s}$ and $\mathbf{w}$.
In this paper, we strive to optimizing the detection performance of one-bit MIMO radar via designing $\mathbf{s}$ and $\mathbf{w}$.  Based on the Neyman-Pearson criterion, we formulate the optimization problem as 
\begin{equation}
	\label{problem}
	\max\limits_{\mathbf{w},\mathbf{s}}~~\mathcal{P}_d~~{\rm s.t.}~~\mathcal{P}_f = a_f,~ {\bf{s}} \in {\cal X}^{{N_t}L \times 1 },
\end{equation}
where $\mathcal{P}_d$ and $\mathcal{P}_f$ are probabilities of detection and false-alarm regarding \eqref{eq6}, and $a_f \in [0,1]$ is a desired false-alarm.

For traditional MIMO radar with infinite-bit quantization, previous studies have shown that the detection performance is positively correlated with the SINR \cite{B11,B12,wu2017transmit}.
In other words, the problem \eqref{problem} is equivalent to a problem of SINR maximization if infinite-bit ADCs are employed.
However, for MIMO radar with one-bit quantization, research to date has not yet determined the detection performance regarding  \eqref{eq6}, i.e., the closed-form expressions of $\mathcal{P}_d$ and $\mathcal{P}_f$.
Therefore, there are two primary aims of this paper: the first is to evaluate the detection performance of one-bit MIMO radar and the second is to implement the joint design.

\section{Detection Performance Evaluation for One-Bit MIMO Radar}
This section seeks to assess the detection performance of one-bit MIMO radar, including the probabilities of false-alarm and detection, by considering a low input SNR/INR scenario.

\subsection{Low Input SNR/INR Assumption}
To derive the detection performance of one-bit MIMO radar explicitly, we have an important assumption, called low input SNR/INR (LIS) assumption.
This assumption shows that before the receiving processing, the input SNR/INR per sample is relatively low, i.e.,
\begin{equation}
\label{eq8}
	 \left| {{h_{1,l}}} \right|^2,\left| {{h_{0,l}}} \right|^2 \ll \sigma^2,~l=1,2,...,N_rL,
\end{equation}
where $ {{h_{1,l}}} $ and $ {{h_{0,l}}} $ are the $l$-th elements of $\mathbf{h}_1$ and $\mathbf{h}_0$, respectively.

It is worth mentioning that the low cost/complexity property of one-bit MIMO radar makes the employment of large bandwidth signals as well as large scale antenna arrays possible, which indicates that rather high processing gains are achievable at the receive end.
Due to the high processing gains, the one-bit MIMO radar will likely operate at rather low SNR/INR values at each receive antenna.
Therefore, the LIS assumption is reasonable in the context of one-bit MIMO radar.

\subsection{Statistical Characteristics of One-Bit Receive Signal  under LIS Assumption}
To evaluate the detection performance, the statistical characteristics of  $\mathbf{y}$ and $z=\mathbf{w}^H\mathbf{y}$ should be derived.
To this end, we consider following proposition.

\textbf{Proposition 1:} Consider a complex variable $x = h + v$, where $h$ is a  constant and $v \sim {\cal C}{\cal N}\left( {0,\sigma _{}^2} \right)$.
Letting $y=\mathcal{Q}(x)$, the following approximation holds up to the first order in $ \epsilon =\frac{h}{\sigma}$, i.e.,
\begin{equation}
\label{pro1-2}
\mathbb{E}\{ y \} =\sqrt{ \frac{4}{\pi\sigma^2} } h + o(\epsilon),~ \mathbb{D}\{ y \} = 2 + o(\epsilon),
\end{equation} 
where $o(\cdot)$ denotes the high order infinitesimal.

\begin{IEEEproof}
	see \textbf{Appendix A}.
\end{IEEEproof}
	
\begin{figure}[!t]
	\centerline{\includegraphics[width=0.4\textwidth]{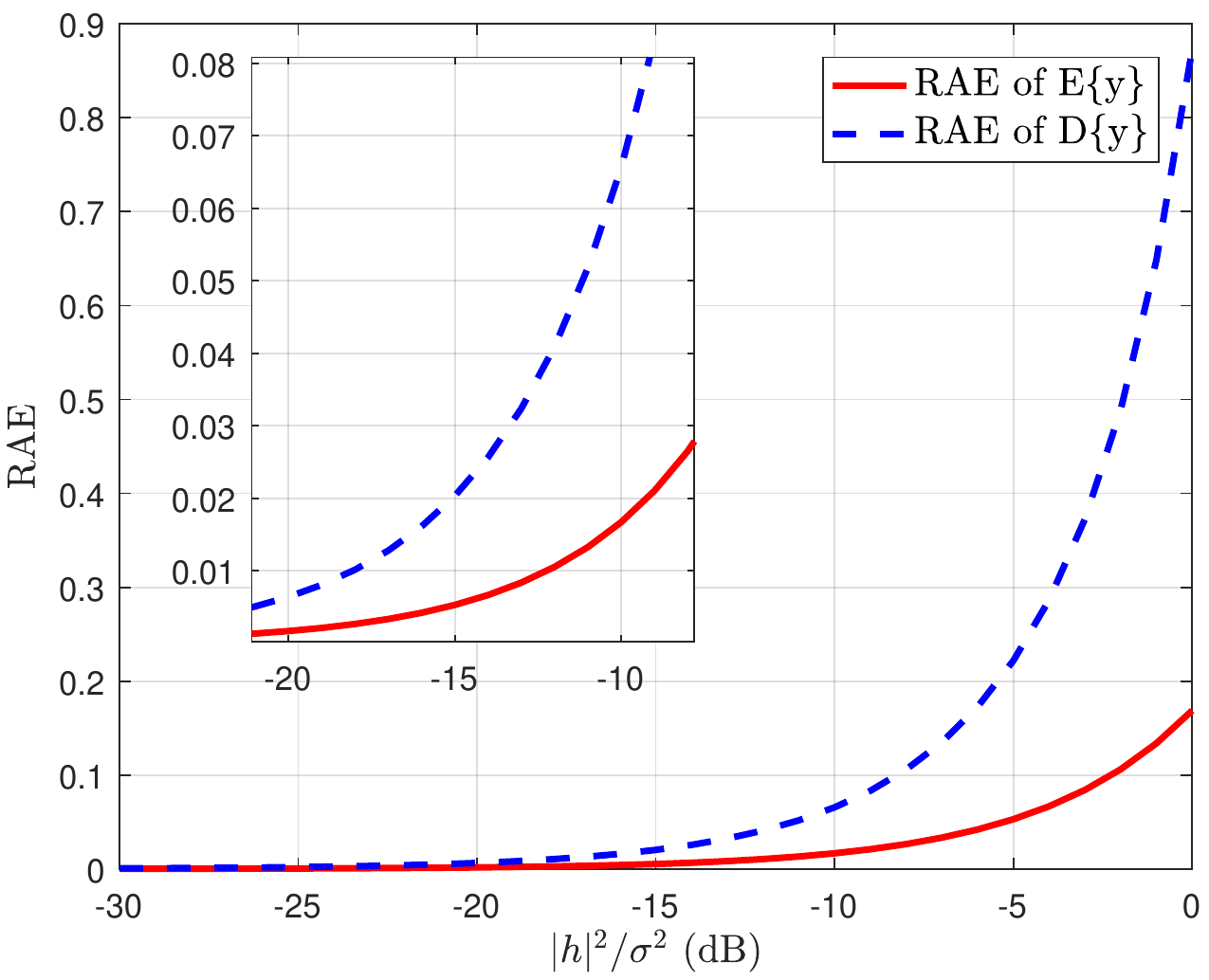}}
	\caption{Relative approximation error (RAE) of $\mathbb{E}\{ y \}$ and $\mathbb{D}\{ y \}$ versus $|h|^2/\sigma^2$, where we set $\Re\{h\}= \Im\{h\}$. }
	\label{fig0}
\end{figure}

Considering different $|h|^2/\sigma^2$, Fig. \ref{fig0} shows the relative approximation error (RAE) of $\mathbb{E}\{ y \}$ and $\mathbb{D}\{ y \}$, i.e., the ratio of approximation error to true value.  
As can be seen,  the RAE decreases with the decreasing of $|h|^2/\sigma^2$.
For both $\mathbb{E}\{ y \}$ and $\mathbb{D}\{ y \}$, their RAEs are less than $0.07$  if $|h|^2/\sigma^2 \leq -10$dB and less than $0.01$ when $|h|^2/\sigma^2 \leq -20$dB.
Hence, a good approximation can be achieved if $|h|^2/\sigma^2$ is small.
	
In the context of one-bit MIMO radar, we note that any element of the one-bit receive signal $\mathbf{y}$ can be regarded as the ``$y$'' in \textbf{Proposition 1}.
Therefore, under the LIS assumption, the \textbf{Proposition 1} can be applied to obtain the statistical characteristics of $\mathbf{y}$.

\subsubsection{Statistical Characteristics of $\mathbf{y}$} 
If $\xi_0,\xi_1,...,\xi_K$ are deterministic,  the elements of $\mathbf{y}$ are independent with each other because the noise $\mathbf{v} \sim \mathcal{CN}(\mathbf{0}_{N_rL},\sigma^2\mathbf{I}_{N_rL})$. 
Therefore, under the LIS assumption, the conditional mean and covariance matrix of  $\mathbf{y}$ can be expressed as
\begin{subequations}
	\label{sta_y}
	\begin{align}
		 & \mathbb{E}\{ \mathbf{y}| \mathcal{H}_i, \xi_0,...,\xi_K\}= \sqrt{\frac{4}{\pi \sigma^2}} \mathbf{h}_i , \\
& \mathbb{D}\{ \mathbf{y}| \mathcal{H}_i, \xi_0,...,\xi_K \}= 2\mathbf{I}_{N_rL},
	\end{align}
\end{subequations}
where $i$ can be $0$ or $1$.

\subsubsection{Statistical Characteristics of $z$} 
Using the results in \eqref{sta_y}, the conditional mean and variance of $z$ can be given by
\begin{subequations}
\begin{align}
	&\mathbb{E}\{ z| \mathcal{H}_i,\xi_0,...,\xi_K  \} = \sqrt{\frac{4}{\pi \sigma^2}}  \mathbf{w}^H\mathbf{h}_i, 
	\\
	 &\mathbb{D}\{ z| \mathcal{H}_i, \xi_0,...,\xi_K   \} =  2\mathbf{w}^H\mathbf{w}.
\end{align}
\end{subequations}
Moreover, with deterministic $\xi_0,\xi_1,...,\xi_K$, we know that $z = \mathbf{w}^H\mathbf{y}$ is a summation of $N_rL$ independent random variables  which share an identical type of distribution.
According to the central-limit theorem (CLT) \cite{billingsley2008probability}, the asymptotical distribution of $z$ is Gaussian.
Therefore, if $\mathbf{h}_i$ satisfies the LIS assumption and the number of receive samples $N_rL$ is large, the conditional  asymptotical probability density function (PDF) of $z$ can be given by
\begin{equation}
p_z( z| \mathcal{H}_i, \xi_0,...,\xi_K ) = \frac{1}{ 2\pi \mathbf{w}^H\mathbf{w}  } {\rm exp} \left(  - \frac{ |z-  \sqrt{\frac{4}{\pi \sigma^2}}  \mathbf{w}^H\mathbf{h}_i|^2 }{ 2 \mathbf{w}^H\mathbf{w} }  \right).
\end{equation}

\color{black}

\subsection{Detection Performance under LIS Assumption}
Having discussed the statistical characteristics of $z$ under the LIS assumption, we are now in a position to investigate the detection performance for one-bit MIMO radar.
Before proceeding, it is important to determine the model of target and interference.
Specifically, we employ a generally accepted model of interference which assumes that $\xi_1,\xi_2,...,\xi_K$ are complex Gaussian variables  $\xi_k \sim \mathcal{CN}(0,\sigma_k^2)$, $k=1,2,\cdots, K$ \cite{B11}.
For target of interest, we consider two cases, where the former refers to a Rayleigh fluctuating target (RFT) with $\xi_0 \sim \mathcal{CN}(0,\sigma_0^2)$ and the latter considers a nonfluctuating target (NFT) with  $\xi_0 = \alpha_0$.
Moreover, we assume these coefficients are independent with each other and also independent with the noise signal.

\subsubsection{Probability of false-alarm}
Letting $\bm{\xi} = [\xi_1,\xi_2,...,\xi_K]^T$, $\beta_k = \sqrt{\frac{4}{\pi \sigma^2}} \mathbf{s}^H \mathbf{A}^H(\theta_{k}) \mathbf{w}$, and $\bm{\beta} = [\beta_1,\beta_2,...,\beta_K]^T$,  we have $\bm{\beta}^H \bm{\xi} =  \sqrt{\frac{4}{\pi \sigma^2}}  \mathbf{w}^H\mathbf{h}_0$. Then the conditional PDF of $z$ under $\mathcal{H}_0$ can be written in a concise form as
\begin{equation}
p_z( z| \mathcal{H}_0, \bm{\xi} ) = \frac{1}{ 2\pi \mathbf{w}^H\mathbf{w}  } {\rm exp} \left(  - \frac{ |z-  \bm{\beta}^H \bm{\xi}|^2 }{ 2 \mathbf{w}^H\mathbf{w} }  \right).
\end{equation}
Moreover, the PDF of $\bm{\xi} $ is denoted as
\begin{equation}
p_{\bm{\xi}}( \bm{\xi} ) = \frac{1}{\pi^K  \det \bm{ \Sigma } } {\rm exp} \left( - \bm{\xi}^H \bm{\Sigma}^{-1} \bm{\xi} \right),
\end{equation} 
where the matrix $\bm{\Sigma} =  {\rm diag}\{\sigma_1^2,\sigma_2^2,...,\sigma_K^2  \} $ and $\det (\cdot)$ is the determinant of a square matrix.	
Then, 
\begin{equation}
\begin{aligned}
p_z( z| \mathcal{H}_0 ) &= \int  p_z( z| \mathcal{H}_0, \bm{\xi} )	p_{\bm{\xi}}( \bm{\xi} ) {\rm d} \bm{\xi} \\
&= \frac{1}{\pi \sigma_{in }^2 } {\rm exp} \left(  - \frac{|z|^2}{ \sigma_{in }^2   }   \right),
\end{aligned}
\end{equation}
where $\sigma_{in }^2 = 2 \mathbf{w}^H\mathbf{w}+ \bm{\beta}^H  \bm{\Sigma} \bm{\beta}$ and the detail derivation  is given in \textbf{Appendix B}.

The  PDF $p_z( z| \mathcal{H}_0 )$ indicates that with the LIS assumption, the $z$ under $\mathcal{H}_0$ is a Gaussian variable with zero mean and variance $\sigma_{in }^2$. 
Thus, under the LIS assumption, $\tilde{z} =|z|$ obeys Rayleigh distribution which gives rise to the closed-form expression of $\mathcal{P}_{f}$ as \cite{Add5}
\begin{equation}
\label{P_f}
\mathcal{P}_{f} = \int_{\mathcal{T}}^{\infty} p_{\tilde{z}}( \tilde{z}| \mathcal{H}_0 ){\rm d}\tilde{z} = {\rm exp} \left( -\frac{ \mathcal{T}^2}{ \sigma^2_{in } }  \right),
\end{equation}
where $ p_{\tilde{z}}( \tilde{z}| \mathcal{H}_0 )$ denotes the conditional PDF of $\tilde{z}$.
Moreover, it is noted that the detection threshold can be set according to a desired false-alarm, i.e., $\mathcal{T} = \sqrt{-\sigma_{in}^2\ln\mathcal{P}_f}$.

Next, we evaluate the probability of detection ($\mathcal{P}_d$) for RFT and NFT, respectively.
\subsubsection{$\mathcal{P}_d$ of RFT}
In the former case, the target is a Rayleigh fluctuating target with $\xi_0 \sim \mathcal{CN}(0,\sigma_0^2)$, in which $\xi_0$ is independent with ${\xi _1},{\xi _2},...,{\xi _K}$ and the system noise. Similar to the derivation of $p_z( z| \mathcal{H}_0 )$,  $p_z( z| \mathcal{H}_1 )$ can be given by
\begin{equation}
p_z( z| \mathcal{H}_1 ) = \frac{1}{\pi (\sigma_0^2|\beta_0|^2 + \sigma_{in}^2   ) } {\rm exp} \left(  - \frac{|z|^2}{ \sigma_0^2|\beta_0|^2 + \sigma_{in}^2       }   \right),
\end{equation}
where $\beta_0 = \sqrt{\frac{4}{\pi \sigma^2}} \mathbf{s}^H \mathbf{A}^H(\theta_{0}) \mathbf{w}$.
Thus, the $z$ under $\mathcal{H}_1$ is also a Gaussian variable with distribution $z\sim \mathcal{CN }(0, \sigma_0^2|\beta_0|^2 + \sigma_{in}^2 )$. Then, the probability of detection is
\begin{equation}
\label{eq15}
\mathcal{P}_{d} = {\rm exp} \left( -\frac{ \mathcal{T}^2}{\sigma_0^2|\beta_0|^2 +  \sigma^2_{in } }  \right) = {\rm exp} \left( \frac{\ln \mathcal{P}_f   }{ 1+   \sigma_0^2|\beta_0|^2/ \sigma_{in}^2}  \right).
\end{equation}

\subsubsection{$\mathcal{P}_d$ of NFT}
For this case, the target reflection coefficient is a constant, i.e., $\xi_0 = \alpha_0$. 
The conditional PDF under $\mathcal{H}_1$ can be denoted as\footnote{The derivation is the same as that of $p_z( z| \mathcal{H}_0 )$, by letting $z=z-\alpha_0\beta_0$.}
\begin{equation}
p_z( z| \mathcal{H}_1 ) = \frac{1}{\pi \sigma_{in}^2    } {\rm exp} \left(  - \frac{|z-\alpha_0\beta_0|^2}{ \sigma_{in}^2       }   \right).
\end{equation}
We observe that $z$ is a Gaussian variable with a nonzero mean $\alpha_0\beta_0$. Thus, $\tilde{z} =|z|$ obeys Rice distribution, and the detection probability of a nonfluctuating target can be written as \cite{Add5}
\begin{equation}
\label{eq17}
\mathcal{P}_d = Q_M( \sqrt{2 |\alpha_0 |^2 | \beta_0 |^2 /\sigma_{in}^2     }  , \sqrt{ - 2\ln \mathcal{P}_f}  ),
\end{equation}
where the function $Q_M(\cdot,\cdot)$ denotes the Marcum function of order 1.

\section{Definition of Quantized SINR}
In the last section, we derive the closed-form expressions of $\mathcal{P}_d$ and $\mathcal{P}_f$ for one-bit MIMO radar under the LIS assumption.
Now, based on the resulting $\mathcal{P}_d$ and $\mathcal{P}_f$, we will introduce the definition of quantized SINR (QSINR).
Similar to the SINR for traditional infinite-bit systems, the QSINR plays a key role affecting the detection performance of one-bit MIMO radar.

\subsection{QSINR with Known Interference Angles}
Observing \eqref{eq15} and \eqref{eq17}, if $\mathcal{P}_f$ is fixed, the $\mathcal{P}_d$ of both the RFT and NFT are monotonically increasing with the increasing of $|\beta_0|^2/ \sigma_{in}^2$. 
Similar to the definition of SINR \cite{Add5}, we can define the QSINR for one-bit MIMO radar under the  LIS assumption as
\begin{equation}
\label{eq18}
{\rm QSINR(\mathbf{w},\mathbf{s})} \triangleq
\begin{cases}
\sigma_0^2   \rho(\mathbf{w},\mathbf{s} ),&\text{RFT},\\
|\alpha_0|^2 \rho(\mathbf{w},\mathbf{s} ),&\text{NFT},
\end{cases}
\end{equation}
where 
\begin{equation}
\begin{aligned}
\rho(\mathbf{w},\mathbf{s} ) = \frac{|\beta_0|^2 }{  \sigma_{in}^2 }  = \frac{2}{ \pi \sigma^2 }  \dfrac{|\mathbf{w}^H \mathbf{A}(\theta_0) \mathbf{s}|^2}{  \mathbf{w}^H \bm{ \Xi }(\mathbf{s}) \mathbf{w} + \mathbf{w}^H\mathbf{w}  }, 
\end{aligned}
\end{equation}
and the matrix $\mathbf{\Xi}\left( {\bf{s}} \right) \in \mathbb{C}^{N_rL \times N_rL} $  is given by
\begin{equation}
\label{Xi25}
\mathbf{\Xi} \left( {\bf{s}} \right) = \sum\limits_{k = 1}^K { \frac{2\sigma_k^2}{\pi\sigma^2 } {\bf{A}}\left( {{\theta _k}} \right){\bf{s}}{{\bf{s}}^H}{{\bf{A}}^H}\left( {{\theta _k}} \right)}. 
\end{equation}
Moreover, since the energy of the transmit waveform is assumed to be 1, i.e., $|| \mathbf{s}||_2^2=1$, $	\rho(\mathbf{w},\mathbf{s} )$ is also equivalent to 
\begin{equation}
\rho(\mathbf{w},\mathbf{s}) =\frac{2}{\pi \sigma^2} \dfrac{  | \mathbf{w}^H \mathbf{A}(\theta_0) \mathbf{s} |^2 }{  \mathbf{s}^H \mathbf{\Phi}(\mathbf{w}) \mathbf{s}   },
\end{equation}
where the matrix
\begin{equation}
\label{Phi27}
{\bf{\Phi }}\left( {\bf{w}} \right) = \sum\limits_{k = 1}^K { \frac{2 \sigma_k^2}{\pi \sigma^2 } {{\bf{A}}^H}\left( {{\theta _k}} \right){\bf{w}}{{\bf{w}}^H}{\bf{A}}\left( {{\theta _k}} \right) + || \mathbf{w} ||^2_2 \mathbf{I}_{N_tL}}.
\end{equation}\color{black}

In addition, it is worth mentioning that the QSINR can be interpreted from the perspective of average output power of $z$. Specifically, under the LIS assumption, the average output power in the absence of target is
\begin{equation}
    \mathbb{E}\{ | \mathbf{w}^H \mathcal{Q}(\mathbf{h}_0+ \mathbf{v})|^2 \} = \mathbb{E}\{ |z|^2| \mathcal{H}_0 \} = \sigma_{in}^2.
\end{equation}
When the target is present,  the average output power of $z$ is
	\begin{equation}
	\begin{aligned}
	\mathbb{E}\{ | \mathbf{w}^H \mathcal{Q}(\mathbf{h}_1+ \mathbf{v})|^2 \} &= \mathbb{E}\{ |z|^2| \mathcal{H}_1 \} \\
	&= \begin{cases}
     \sigma_0^2|\beta_0|^2 + \sigma_{in}^2, & \text{RFT}, \\
     |\alpha_0\beta_0|^2+ \sigma_{in}^2, & \text{NFT}.
    \end{cases}
	\end{aligned}
\end{equation}
Thus, the QSINR can also be expressed as
\begin{equation}
\label{QSINR_prac}
    {\rm QSINR(\mathbf{w},\mathbf{s})} = \frac{\mathbb{E}\{ | \mathbf{w}^H \mathcal{Q}(\mathbf{h}_1+ \mathbf{v})|^2 \}}{\mathbb{E}\{ | \mathbf{w}^H \mathcal{Q}(\mathbf{h}_0+ \mathbf{v})|^2 \}}-1,
\end{equation}
where the first term refers to the ratio of average output power in the presence of target to that in the absence of target, and the second term ``$-1$'' is employed to remove the interference and noise part in $\mathbb{E}\{ | \mathbf{w}^H \mathcal{Q}(\mathbf{h}_1+ \mathbf{v})|^2 \}$.

\subsection{QSINR under Stochastic Case}	
The {\color{black} QSINR given by \eqref{eq18} } requires full knowledge of the interference directions  $\theta_1,\theta_2,...,\theta_K$.
Nevertheless, in practical applications, the exact location of the interference might be unknown \cite{B12}.
For such scenarios, it is reasonable to assume that $\theta_1,\theta_2,...,\theta_K$ are random variables. 
For convenience, we consider  the normalized angle $\omega=\sin \theta\in[0,1]$ instead of $\theta$. We assume that
the normalized angle of interference $\omega_k=\sin \theta_k$ ($k=1,2,...,K$) obeys the uniform distribution with known mean $ \varpi_k  $ and uncertainty bound $\delta_k $, i.e.,
\begin{equation}
\omega_k \sim \mathcal{U}( \varpi_k  -\delta_k, \varpi_k  +\delta_k ),~ k=1,2,...,K.
\end{equation}

Consider the matrix form of $\mathbf{w}$ and $\mathbf{s}$, i.e., $\mathbf{W}\in \mathbb{C}^{N_r \times L}$ and {\color{black} $\mathbf{S}\in \mathbb{C}^{N_t \times L}$}, respectively.
For the case of interference with angle uncertainty, $\mathbf{\Xi}\left( {\bf{s}} \right) $ and ${\bf{\Phi }}\left( {\bf{w}} \right)$  can be written as
\begin{subequations}
	\label{Xi&Phi}
	\begin{align}
	&\mathbf{\Xi} ({\bf{s}}) = \sum\limits_{k = 1}^K { {\color{black} \frac{2\sigma_k^2}{\pi\sigma^2 }  } {\left( {{\bf{S}}_{}^T \otimes {{\bf{I}}_{{N_r}}}} \right){\mathbf{C}(\omega_k)}{{\left( {{\bf{S}}_{}^T \otimes {{\bf{I}}_{{N_r}}}} \right)}^H}}}, \\
	&{\bf{\Phi }} ({\bf{w}}) =\sum\limits_{k = 1}^K {{\color{black} \frac{2\sigma_k^2}{\pi\sigma^2 }  }\left( {{\bf{W}}_{}^T \otimes {{\bf{I}}_{{N_t}}}} \right){\mathbf{D}(\omega_k)}{\left( {{\bf{W}}_{}^T \otimes {{\bf{I}}_{{N_t}}}} \right)^H}} \\
	&~~~~~~~~~~~	+ || \mathbf{w} ||^2_2 \mathbf{I}_{N_tL},
	\end{align}
\end{subequations}
\color{black}
where the matrices $\mathbf{C}(\omega_k),~\mathbf{D}(\omega_k) \in \mathbb{C}^{N_tN_r \times N_tN_r}$. 	For the $(m,n)$-th ($m,n=1,2,...,N_t$) block of  $\mathbf{C}(\omega_k)$, the $(p,q)$-th ($p,q=1,2,...,N_r$) element is given by
\begin{equation}
\mathbf{C}_{(m,n)}^{(p,q)}(\omega_k) = \frac{e^{-\jmath\pi \tilde{d}_{(m,n)}^{(p,q)} \varpi_k }}{N_tN_r} \text{sinc} ( \tilde{d}_{(m,n)}^{(p,q)} \delta_k ),
\end{equation}
where $\tilde{d}_{(m,n)}^{(p,q)} = 2(d_{t,m}-d_{t,n} + d_{r,p} -d_{r,q})/\lambda $ and the function $\text{sinc}(x) \triangleq \frac{\sin \pi x  }{\pi x}$. 
For the $(m,n)$-th ($m,n=1,2,...,N_r$) block of  $\mathbf{D}(\omega_k)$, the $(p,q)$-th ($p,q=1,2,...,N_t$) element is given by
\begin{equation}
\mathbf{D}_{(m,n)}^{(p,q)}(\omega_k) = \frac{e^{\jmath\pi \tilde{d}^{(m,n)}_{(p,q)} \varpi_k }}{N_tN_r} \text{sinc} ( \tilde{d}^{(m,n)}_{(p,q)} \delta_k ).
\end{equation}
The detail derivation of $\mathbf{C}(\omega_k)$ and $\mathbf{D}(\omega_k)$ is given in \textbf{Appendix C}.

\color{black}
Based on the above analyses, the detection performance of one-bit MIMO radar under the LIS assumption is determined by the QSINR.
Therefore, to achieve the best detection performance, the QSINR should be maximized.
Moreover, for both the RFT and NFT cases, maximizing QSINR is essentially maximizing $\rho(\mathbf{w},\mathbf{s})$, which requires a joint design of $\mathbf{w}$ and $\mathbf{s}$. Before tackling this joint design problem, it is worth investigating the performance gap between the one-bit and infinite-bit MIMO radars, which is provided in the next subsection. 
\color{black}

\subsection{\textcolor{black}{SINR Comparison between One-Bit and Infinite-Bit MIMO Radars}}
Employing the one-bit converters can significantly reduce the system complexity of a MIMO radar at the cost of inevitable performance degeneration to some degree.
Therefore, it is necessary to compare the performance between one-bit and traditional infinite-bit MIMO radars.

\subsubsection{Loss of using one-bit ADCs}  
Observing the QSINR defined by \eqref{eq18} and the SINR for MIMO radar with infinite-bit ADCs \cite{B11,B12,wu2017transmit}, the difference is the constant term $2/\pi$. 
Therefore, in the low SNR/INR regime, the loss of the output QSINR caused by introducing one-bit ADCs is about $2/\pi \approx1.96$dB with reference to the traditional MIMO radar with infinite-bit ADCs.

\subsubsection{Loss of using one-bit DACs} 
In the noise-only case, the loss of using one-bit DACs can be evaluated from the transmit beampattern point of view.
The transmit beampattern, which indicates the transmit power distribution in spatial domain, is defined as  \cite{A5}
\begin{equation}
\begin{aligned}
		F(\theta_{0}) & \triangleq || \mathbf{a}^T_t(\theta_{0})\mathbf{S}||_2^2 \\
		& = \frac{1}{N_t^2L} \sum_{l=1}^{L} \left|  \sum_{k=1}^{N_t}  e^{\jmath\varphi_{k,l}-\jmath2\pi d_{t,k}\sin \theta_0 /\lambda}    \right|^2,
\end{aligned}
\end{equation}
where $\varphi_{k,l}$ denotes the phase of $\mathbf{S}(k,l)$, $k=1,2,...,N_t,~l=1,2,...,L$.

For traditional MIMO radar with infinite-bit DACs, we can set $\varphi_{k,l}= 2\pi d_{t,k} \sin \theta_{0}/\lambda$. Then, $F(\theta_0)=1$, and this is the ideal case without transmit loss.
However, for MIMO radar with one-bit DACs, we only have four possible $\varphi_{k,l}$ to choose, i.e., $\varphi_{k,l} \in \{ \frac{\pi}{4}, \frac{3\pi}{4},\frac{5\pi}{4},\frac{7\pi}{4}\}$.
Therefore, the phase of the transmit waveform can not exactly match with that of the target steering vector.
In fact, we can set
\begin{equation}
\label{varphi_solution}
\varphi_{k,l} = \begin{cases}
\frac{1}{4}\pi,    		\mod(\phi_{k,l}, 2\pi)  \in [0,\frac{1}{2}\pi), \\
\frac{3}{4}\pi, 	\mod(\phi_{k,l}, 2\pi)  \in [\frac{1}{2}\pi,\pi), \\
\frac{5}{4}\pi, 		\mod(\phi_{k,l}, 2\pi)  \in [\pi,\frac{3}{2}\pi), \\
\frac{7}{4}\pi, \mod(\phi_{k,l}, 2\pi)  \in [\frac{3}{2}\pi,2\pi), \\
\end{cases}
\end{equation}
where $\phi_{k,l}=2\pi d_{t,k}\sin \theta_0/\lambda+\frac{\pi}{4}$ and $\text{mod}(x,y)$ is the remainder after division of $x$ by $y$.

When $\varphi_{k,l}$ satisfying \eqref{varphi_solution}, we know all the residual phases $\Delta \varphi_{k,l} \in (0, \frac{1}{2}\pi]$, where $\Delta \varphi_{k,l} = \varphi_{k,l}-2\pi d_{t,k} \sin \theta_{0}/\lambda$. 
Note that, the worst case happens when half of these residual phases satisfying  $\Delta \varphi_{k,l}\rightarrow 0$ and the other half satisfying $\Delta \varphi_{k,l}\rightarrow 0.5\pi$, which results in $F(\theta_{0})\rightarrow 0.5$.
Therefore, in the noise-only case, the QSINR loss caused by using one-bit DACs is less than 3dB if $\varphi_{k,l}$ satisfies \eqref{varphi_solution}.

\color{black}

\section{Solution to the Joint Design}
In the last section, we show that under the LIS assumption, the maximization of  QSINR can achieve the best detection performance for one-bit MIMO radar.
Thus, the explicit form of the joint optimization problem \eqref{problem} can be given by
\begin{equation} \label{problem1}
\max\limits_{\mathbf{w},\mathbf{s}}~~ \rho(\mathbf{w},\mathbf{s}) ~~ \text{s.t.}~~{\bf{s}} \in {\cal X}^{{N_t}L \times 1 }.  
\end{equation}
It is noted that the binary constraint and the couple of $\mathbf{w}$ and $\mathbf{s}$ in $\rho(\mathbf{w},\mathbf{s})$ are the main difficulties in solving the problem.

In this paper, we shall propose an alternatin\emph{g} wavefo\emph{r}m and filt\emph{e}r d\emph{e}sign for QSINR maximiza\emph{t}ion (GREET). 
The core idea of the GREET lies in the alternating optimization (AltOpt) which properly separates the coupled variables and sequentially optimizes them. 
The AltOpt has been proved to be an efficient manner to deal with the problem with a objective like $\rho(\mathbf{w},\mathbf{s})$ \cite{6104176,6472022,B10,B11,B12,wu2017transmit}.  
Interestingly, \cite{wu2017transmit} reveals that the AltOpt can be interpreted from the perspective of majorization-minimization \cite{sun2016majorization}.
Specifically, in the GREET framework, we first optimize the receive filter $\bf w$ with a fixed waveform $\bf s$, and then we use the ADMM method,  which is able to converge to a KKT point, to optimize $\bf s$ with a given $\bf w$.

\color{black}

\subsection{Optimizing the receive filter $\bf w$ with a fixed $\mathbf{s}$}
Firstly, for a fixed  $\mathbf{s}$, the problem \eqref{problem1} w.r.t. $\mathbf{w}$ is simplified to
\begin{equation} 
\label{subproblem_w}
	\max\limits_{\mathbf{w}}~~  \frac{ {{{\left| {{{\bf{w}}^H}{\bf{A}}\left( {{\theta _0}} \right){\bf{s}}} \right|}^2}}}{{{{\bf{w}}^H}\mathbf{\Xi} \left( {\bf{s}} \right){\bf{w}} + {{\bf{w}}^H}{\bf{w}}}}.
\end{equation}
It is easy to know that the problem \eqref{subproblem_w} is equivalent to the well-known MVDR problem \cite{add3},  whose closed-form solution is given by 
\begin{equation}
	{\bf{w}} = \frac{{{{\left( {{\bf{\Xi }}\left( {\bf{s}} \right) + {\bf{I}}_{N_rL}} \right)}^{ - 1}}{\bf{A}}\left( {{\theta _0}} \right){\bf{s}}}}{{{{\bf{s}}^H}{{\bf{A}}^H}\left( {{\theta _0}} \right){{\left( {{\bf{\Xi }}\left( {\bf{s}} \right) + {\bf{I}}_{N_rL}} \right)}^{ - 1}}{\bf{A}}\left( {{\theta _0}} \right){\bf{s}}}}.
\end{equation}

\subsection{Optimizing the transmit waveform  $\mathbf{s}$  with a fixed $\mathbf{w}$}
For a given $\mathbf{w}$, the problem \eqref{problem1} is equivalent to
\begin{equation} \label{subproblem_s}
	\min\limits_{\mathbf{s}}~~  \frac{ \mathbf{s}^H \mathbf{\Phi}(\mathbf{w}) \mathbf{s} }{ \mathbf{s}^H  \mathbf{\Gamma}(\mathbf{w}) \mathbf{s} }  
	~~\text{s.t.}~~{\bf{s}} \in {\cal X}^{{N_t}L \times 1  },  
\end{equation}
where  $\mathbf{\Gamma}(\mathbf{w}) = \mathbf{A}^H(\theta_0)\mathbf{w}\mathbf{w}^H\mathbf{A}(\theta_0)$.

{\color{black}
Since both $\mathbf{\Phi}(\mathbf{w})$ and $\mathbf{\Gamma}(\mathbf{w})$ are conjugate symmetric matrices, the subproblem \eqref{subproblem_s} can be reformulated in a real-valued form as \cite{chi2017convex}
\begin{equation} \label{subproblem_sr}
\min\limits_{\mathbf{\tilde{s}}}~~  \frac{ \mathbf{\tilde{s}}^T \mathbf{\tilde{\Phi}}(\mathbf{w})  \mathbf{\tilde{s}} }{ \mathbf{\tilde{s}}^T \mathbf{\tilde{\Gamma}}(\mathbf{w}) \mathbf{\tilde{s}} }~~  \text{s.t.}~~\mathbf{\tilde{s}} \in \dfrac{1}{{\sqrt {2{N_t}L} }}{\{ 1,-1\} ^{2{N_t}L  \times 1}}, 
\end{equation}
where  }
\begin{subequations}
	\begin{align}
&{\bf{\tilde s}} = \left[ {\begin{array}{*{20}{c}}
	{\Re \left\{ {\bf{s}} \right\}}\\
	{\Im \left\{ {\bf{s}} \right\}}
	\end{array}} \right],  \\
&{\bf{\tilde \Gamma }}\left( {\bf{w}} \right) = \left[ {\begin{array}{*{20}{c}}
	{\Re \left\{ {{\bf{\Gamma }}\left( {\bf{w}} \right)} \right\}}&{ - \Im \left\{ {{\bf{\Gamma }}\left( {\bf{w}} \right)} \right\}}\\
	{\Im \left\{ {{\bf{\Gamma }}\left( {\bf{w}} \right)} \right\}}&{\Re \left\{ {{\bf{\Gamma }}\left( {\bf{w}} \right)} \right\}}
	\end{array}} \right], \label{21b}\\
&{\bf{\tilde \Phi }}\left( {\bf{w}} \right) = \left[ {\begin{array}{*{20}{c}}
	{\Re \left\{ {{\bf{\Phi }}\left( {\bf{w}} \right)} \right\}}&{ - \Im \left\{ {{\bf{\Phi }}\left( {\bf{w}} \right)} \right\}}\\
	{\Im \left\{ {{\bf{\Phi }}\left( {\bf{w}} \right)} \right\}}&{\Re \left\{ {{\bf{\Phi }}\left( {\bf{w}} \right)} \right\}}
	\end{array}} \right]. 
	\end{align}
\end{subequations}

{\color{black}  The problem \eqref{subproblem_sr} is composed of a fractional quadratic objective and a binary constraint, which is nonconvex and hard to be directly tackled. 
Potential methods  to deal with such a problem include the semidefinite relaxation (SDR) method \cite{Add1} and the block coordinate descent (BCD)   method \cite{Add2}. 
Nevertheless, as shall be shown later in our simulations, the two methods cannot always ensure satisfactory performance. 
Moreover, they may suffer from prohibitive computational burdens, especially in large  $ N_tL $ case.
To this end, a more efficient algorithm based on the ADMM method is developed to tackle the problem.}

{\color{black} To facilitate tackling the optimization problem,  we write the binary constraint in \eqref{subproblem_sr} as the following equivalent form}
\begin{equation}
\label{eq29}
		 \mathbf{\tilde{s}}^T\mathbf{\tilde{s}}=1,~-\dfrac{1}{{\sqrt {2{N_t}L} }} \mathbf{1}_{2N_tL}    \leq  \mathbf{\tilde{s}} \leq \dfrac{1}{{\sqrt {2{N_t}L} }} \mathbf{1}_{2N_tL},
\end{equation}
{\color{black} The proof of the equivalence is explicit. Specifically, the linear constraint in \eqref{eq29} shows that $|\tilde{s_k}|^2 \leq \frac{1}{2N_tL}$, $k=1,2,...,2N_tL$, where $\tilde{s}_k$ is the $k$-th entry of $\mathbf{\tilde{s}}$. 
	Then, we have 
\begin{equation}
	\mathbf{\tilde{s}}^T\mathbf{\tilde{s}}= \sum\limits_{k=1}^{2N_tL} |\tilde{s_k}|^2 \leq 1.
\end{equation}
To ensure $\mathbf{\tilde{s}}^T\mathbf{\tilde{s}}=1$, it is obvious that all $\tilde{s}_k$ must satisfy $|\tilde{s}_k|=\frac{1}{\sqrt{2N_tL}}$. Hence, \eqref{eq29} is equivalent to the binary constraint.
 }

Next, to implement the ADMM framework \cite{BS1} to solve  the problem \eqref{subproblem_sr}, we introduce two variables $\mathbf{t}, \mathbf{r} \in \mathbb{R}^{2N_tL\times1}$ to reformulate the problem \eqref{subproblem_sr} as
\begin{subequations} \label{subproblem_sr1}
	\begin{align}
	\min\limits_{\mathbf{\tilde{s}},\mathbf{t},\mathbf{r}}~~   &\frac{ \mathbf{t}^T \mathbf{\tilde{\Phi}}(\mathbf{w})  \mathbf{t} }{ \mathbf{r}^T \mathbf{\tilde{\Gamma}}(\mathbf{w}) \mathbf{r} }\\
	\text{s.t. }~~~&\mathbf{t}=\mathbf{\tilde{s}},~\mathbf{r}=\mathbf{\tilde{s}},~ \mathbf{t}^T\mathbf{t}=1, \\
	&-\dfrac{1}{{\sqrt {2{N_t}L} }} \mathbf{1}_{2N_tL}   \leq  \mathbf{\tilde{s}} \leq \dfrac{1}{{\sqrt {2{N_t}L} }} \mathbf{1}_{2N_tL} .
	\end{align}
\end{subequations}
Then, the scaled augmented Lagrangian function of \eqref{subproblem_sr1}  is given by
\begin{equation}
\begin{aligned}
	{{\cal L}_{{\rho _1},{\rho _2}}}\left( {{\bf{\tilde s}},{\bf{t}},{\bf{r}},{{\bf{u}}_1},{{\bf{u}}_2}} \right){\rm{ =   }}&
\frac{{{{{\bf{t}}}^T}{\bf{\tilde \Phi }}\left( {\bf{w}} \right){\bf{t}}}}{{{{\bf{r}}^T}{\bf{\tilde \Gamma }}\left( {\bf{w}} \right){\bf{r}}}} + \frac{{{\rho _1}}}{2}\left\| {{\bf{t}} - {\bf{\tilde s}} + {{\bf{u}}_1}} \right\|_2^2 \\ 	&+ \frac{{{\rho _2}}}{2}\left\| {{\bf{r}} - {\bf{\tilde s}} + {{\bf{u}}_2}} \right\|_2^2,
\end{aligned}
\end{equation}
where $\mathbf{u}_1,\mathbf{u}_2 \in \mathbb{R}^{2N_tL \times 1}$ are dual variables, and $\rho_1,\rho_2>0$ are penalty variables, which place  penalties on the violations of
primal feasibilities $ \mathbf{t}=\mathbf{\tilde{s}}$ and $ \mathbf{r}=\mathbf{\tilde{s}}$.

Under the ADMM framework, the following iterations need to be performed sequentially:
\begin{subequations}
	\label{eq26}
	\begin{align}
		&{\bf{\tilde s}} \leftarrow \mathop {\min }\limits_{{\bf{\tilde s}}}  {{\cal L}_{{\rho _1},{\rho _2}}}\left( {{\bf{\tilde s}},{\bf{t}},{\bf{r}},{{\bf{u}}_1},{{\bf{u}}_2}} \right){\rm{, s}}{\rm{.t}}{\rm{. }}  ~\mathbf{\tilde{s}} \in \mathbb{D} \label{26a}, \\
		&{\bf{t}} \leftarrow \mathop {\min }\limits_{\bf{t}}  {{\cal L}_{{\rho _1},{\rho _2}}}\left( {{\bf{\tilde s}},{\bf{t}},{\bf{r}},{{\bf{u}}_1},{{\bf{u}}_2}} \right){\rm{, s}}{\rm{.t}}{\rm{.  }}~{{\bf{t}}^H}{\bf{t}} = 1 \label{26b},  \\
		&{\bf{r}} \leftarrow \mathop {\min }\limits_{\bf{r}}  {{\cal L}_{{\rho _1},{\rho _2}}}\left( {{\bf{\tilde s}},{\bf{t}},{\bf{r}},{{\bf{u}}_1},{{\bf{u}}_2}} \right)  \label{26c}, \\
		&{{\bf{u}}_1} \leftarrow {{\bf{u}}_1} + \left( {{\bf{t}} - {\bf{s}}} \right),{{\bf{u}}_2} \leftarrow {{\bf{u}}_2} + \left( {{\bf{r}} - {\bf{s}}} \right),
	\end{align}
\end{subequations}
where $ \mathbb{D} = \{\mathbf{\tilde{s}}| - \frac{1}{{\sqrt {2{N_t}L} }}{\mathbf{1}_{2N_tL}} \le {\bf{\tilde s}} \le \frac{1}{{\sqrt {2{N_t}L} }}{\mathbf{1}_{2N_tL}}   \} $.

In the sequel, the solutions to the subproblems \eqref{26a}-\eqref{26c} are presented.
\subsubsection{Update of $\mathbf{\tilde{s}}$}
Omitting the constant term of the objective in the subproblem \eqref{26a}, the update of $\mathbf{\tilde{s}}$ needs to solve following problem:
\begin{subequations} \label{subproblem_srs}
	\begin{align}
	&\min\limits_{\mathbf{\tilde{s}}}~~  \frac{\rho_1+\rho_2}{2} \mathbf{\tilde{s}}^T\mathbf{\tilde{s}} - \mathbf{\tilde{s}}^T \mathbf{b}   \\
	&\text{s.t. }~~~-\dfrac{1}{{\sqrt {2{N_t}L} }} \mathbf{1}_{2N_tL}   \leq  \mathbf{\tilde{s}} \leq \dfrac{1}{{\sqrt {2{N_t}L} }} \mathbf{1}_{2N_tL}, 
	\end{align}
\end{subequations}
where $\mathbf{b} = \rho_1(\mathbf{t}+\mathbf{u}_1) + \rho_2(\mathbf{r}+\mathbf{u}_2)$.
It is easy to notice that the problem \eqref{subproblem_srs} is a convex problem, and its closed-form solution  can be obtained by the KKT condition  \cite{BS2},  as  
\begin{equation}
\label{update_s}
	\tilde{s}_k =
	\begin{cases}
	 \sqrt{\dfrac{1}{2N_tL}}, & \dfrac{b_k}{\rho_1+\rho_2}\geq \sqrt{\dfrac{1}{2N_tL}}, \\
	 \dfrac{b_k}{\rho_1+\rho_2}, &  -\sqrt{\dfrac{1}{2N_tL}} < \dfrac{b_k}{\rho_1+\rho_2} < \sqrt{\dfrac{1}{2N_tL}}, \\
	 -\sqrt{\dfrac{1}{2N_tL}}, & \dfrac{b_k}{\rho_1+\rho_2} \leq -\sqrt{\dfrac{1}{2N_tL}} ,	 
	\end{cases}
\end{equation}
where  $b_k$ is the $k$-th entry of $\mathbf{b}$.


\subsubsection{Update of $\mathbf{t}$}
 The subproblem w.r.t. $\mathbf{t}$ is given by
\begin{subequations} 
	\label{ADMM_t1}
	\begin{align} 
	&\min\limits_{\mathbf{t}}~~   \frac{ \mathbf{t}^T \mathbf{\tilde{\Phi}}(\mathbf{w})  \mathbf{t} }{ \mathbf{r}^T \mathbf{\tilde{\Gamma}}(\mathbf{w}) \mathbf{r} } +\frac{ \rho_1}{2} ||\mathbf{\mathbf{t}-\tilde{s}} +\mathbf{u}_1 ||_2^2\\
	&\text{s.t. }~~~~ \mathbf{t}^T\mathbf{t}=1. 
	\end{align}
\end{subequations}
The above problem is nonconvex due to the quadratic equality constraint. 
To solve it, let $\mathbf{\tilde{\Phi}}(\mathbf{w}) = {\bf{{ P}\Pi }}{{\bf{P}}^T}$ denote the eigenvalue decomposition (EVD) of $\mathbf{\tilde{\Phi}}(\mathbf{w})$, where $ \mathbf{\Pi} $ is a diagonal matrix whose diagonal elements are the eigenvalues of $\mathbf{\tilde{\Phi}}(\mathbf{w})$, i.e., $\gamma_1\geq\gamma_2\geq...\geq\gamma_{2N_tL}$, and $\mathbf{P}$ is a $2N_tL\times 2N_tL$ matrix whose columns are the corresponding right eigenvectors.

Then, using the variable substitution, i.e., $\mathbf{\tilde{t}} = \mathbf{P}^T \mathbf{t}$ and $\mathbf{g}=\mathbf{P}^T (\mathbf{\tilde{s}} - \mathbf{u}_1) $, the optimization problem is reformulated as
\begin{subequations} 
	\label{ADMM_t2}
	\begin{align} 
	&\min\limits_{\mathbf{\tilde{t}}}~~  \eta  \mathbf{\tilde{t}}^T \mathbf{\Pi}\mathbf{\tilde{t}} +\frac{ \rho_1}{2} ||\mathbf{\tilde{t}} -\mathbf{g} ||_2^2\\
	&\text{s.t. }~~~~ \mathbf{\tilde{t}}^T\mathbf{\tilde{t}}=1, 
	\end{align}
\end{subequations}
where $\eta = \frac{ 1 }{ \mathbf{r}^T \mathbf{\tilde{\Gamma}}(\mathbf{w}) \mathbf{r} } $.

The Lagrange function of the problem \eqref{ADMM_t2} can be denoted as
\begin{equation}
	\mathcal{L}(\mathbf{\tilde{t}}, \nu) = \eta  \mathbf{\tilde{t}}^T \mathbf{\Pi}\mathbf{\tilde{t}} +\frac{ \rho_1}{2} ||\mathbf{\tilde{t}} -\mathbf{g} ||_2^2 + \nu(\mathbf{\tilde{t}}^T\mathbf{\tilde{t}}-1 ),
\end{equation}
where $\nu$ is the Lagrange multiplier. According to the KKT condition, we obtain the  optimal condition of the problem \eqref{ADMM_t2}, as
\begin{equation}
2 \eta \mathbf{\Pi} \mathbf{\tilde{t}} + (\rho_1 +2\nu)\mathbf{\tilde{t}}  = \rho_1 \mathbf{g},~		\mathbf{\tilde{t}}^T\mathbf{\tilde{t}}=1, 
\end{equation}
After some mathematical operations, we obtain that
\begin{equation}
\label{eq37}
	\tilde{t}_k = \frac{\rho_1g_k}{2\eta \gamma_k + \rho_1 + 2\nu },~k=1,2,...,2N_tL,
\end{equation}
and
\begin{equation}
\label{BS1}
\sum_{k=1}^{2N_tL} \frac{\rho_1^2g_k^2}{(2\eta \gamma_k + \rho_1 + 2\nu )^2}=1,
\end{equation}
where $\tilde{t}_k $ and $g_k$ are the $k$-th element of $\mathbf{\tilde{t}}$ and $\mathbf{g}$, respectively.

Define function $ f(\nu) \triangleq \sum_{k=1}^{2N_tL} \frac{\rho_1^2g_k^2}{(2\eta \gamma_k + \rho_1 + 2\nu )^2}$. Since $\eta >0$, $\rho_1>0$, and $\gamma_k>0$ for all $k=1,2,...,2N_tL$, $f(\nu)$ is monotonically increasing for $\nu \in (-\infty, -\eta \gamma_{1}-0.5\rho_1)$ and monotonically decreasing for  $\nu \in ( -\eta \gamma_{2N_tL}-0.5\rho_1, \infty)$.
Moreover, when $|\nu|\rightarrow \infty$,  $f(\nu)\rightarrow0$, and when $\nu \rightarrow -\eta \gamma_{1}-0.5\rho_1$ (or $-\eta \gamma_{2N_tL}-0.5\rho_1$), $f(\nu)$ goes to infinity.
Thus, $ f(\nu) =1$ has a unique solution in $(-\infty, -\eta \gamma_{1}-0.5\rho_1)$ or in $( -\eta \gamma_{2N_tL}-0.5\rho_1, \infty)$, and the solution can be numerically obtained via the bisection method (BSM) \cite{BS2}.

Finally, when getting $\nu$, the optimal $\mathbf{\tilde{t}}$ can be computed by \eqref{eq37}, and the final solution to the problem \eqref{ADMM_t1} is $\mathbf{t}= \mathbf{P}\mathbf{\tilde{t}}$.

\subsubsection{Update of $\mathbf{r}$}
To update $\mathbf{r}$, we need to solve following unconstrained problem:
\begin{equation}
	\min\limits_{\mathbf{r}}~~   \frac{ \mathbf{t}^T \mathbf{\tilde{\Phi}}(\mathbf{w})  \mathbf{t} }{ \mathbf{r}^T \mathbf{\tilde{\Gamma}}(\mathbf{w}) \mathbf{r} } +\frac{ \rho_2}{2} ||\mathbf{\mathbf{r}-\tilde{s}} +\mathbf{u}_2 ||_2^2.
\end{equation}
To attain its optimal solution, let ${\bf{\tilde \Gamma }}\left( {\bf{w}} \right) = {\bf{U\Lambda }}{{\bf{U}}^T}$ be the EVD of ${\bf{\tilde \Gamma }}$, where $\mathbf{U} \in \mathbb{R}^{2N_tL \times 2N_tL}$ is the eigenmatrix and $\mathbf{\Lambda} = {\rm diag}\{\lambda_1,\lambda_2,...,\lambda_{2N_tL} \}$ with ${\lambda _1} \ge {\lambda _2} \ge ... \ge {\lambda _{2{N_t}L}}$ being the eigenvalues of ${\bf{\tilde \Gamma }}\left( {\bf{w}} \right)$.

Letting $\mathbf{\tilde{r}}= \mathbf{U}^T\mathbf{r}$ and $\mathbf{q} = \mathbf{U}^T(\mathbf{\tilde{s}} -\mathbf{u}_2 )$, the above problem is equivalent to
\begin{equation}
\min\limits_{\mathbf{\tilde{r}}}~~   \frac{ \mathbf{t}^T \mathbf{\tilde{\Phi}}(\mathbf{w})  \mathbf{t} }{ \mathbf{\tilde{r}}^T \mathbf{\Lambda} \mathbf{\tilde{r}} } +\frac{ \rho_2}{2} ||\mathbf{\tilde{r}} -\mathbf{q} ||_2^2.
\end{equation}

Since the rank of   $\mathbf{ {\Gamma}}(\mathbf{w})$ in \eqref{subproblem_s} is 1, we can infer that the rank of $\mathbf{ \tilde{\Gamma}}(\mathbf{w})$ is 2 from the definition of $\mathbf{\tilde{\Gamma}}(\mathbf{w})$ in \eqref{21b},   and the corresponding eigenvalues are equal, i.e.,
\begin{equation}
	\lambda_1=\lambda_2 > 0, ~\lambda_k=0,~k=3,4,..,2N_tL.
\end{equation}
Thus, for $k=3,4,..,2N_tL$, the optimal $\tilde{r}_k=q_k$, where $\tilde{r}_k$ and $q_k$ are the $k$-th elements of $\mathbf{\tilde{r}}$ and $\mathbf{q}$.

To obtain the optimal $\tilde{r}_1$ and $\tilde{r}_2$, we need to solve following problem:
\begin{equation}
\min\limits_{\tilde{r}_1,\tilde{r}_2}~~   \frac{ \mathbf{t}^T \mathbf{\tilde{\Phi}}(\mathbf{w})  \mathbf{t} }{ \lambda_1\tilde{r}_1^2 + \lambda_2\tilde{r}_2^2 } +\frac{ \rho_2}{2} \left[  (\tilde{r}_1-q_1)^2 + (\tilde{r}_2-q_2)^2 \right].
\end{equation}
Let $\tilde \lambda  = {\lambda _1} = {\lambda _2}$ and $p = 2\frac{{{{{\bf{t}}}^T}{\bf{\tilde \Phi }}\left( {\bf{w}} \right){\bf{t}}}}{{\tilde \lambda {\rho _2}}}.$
A concise form of the above problem is given by
\begin{equation}
\label{ADMM_r3}
\min\limits_{\tilde{r}_1,\tilde{r}_2}~~   \frac{ p}{ \tilde{r}_1^2 + \tilde{r}_2^2 } +  (\tilde{r}_1-q_1)^2 + (\tilde{r}_2-q_2)^2 .
\end{equation}

Consider function
\begin{equation}
	\tilde{f}(\tilde{r}_1, \tilde{r}_2) \triangleq  \frac{ p}{ \tilde{r}_1^2 + \tilde{r}_2^2 } +  (\tilde{r}_1-q_1)^2 + (\tilde{r}_2-q_2)^2, 
\end{equation}
then, the optimal condition can be given by
\begin{subequations}
	\label{eq46}
	\begin{align}
			\dfrac{{\partial \tilde{f}\left( {{{\tilde r}_1},{{\tilde r}_2}} \right)}}{{\partial {{\tilde r}_1}}} = \dfrac{{ - 2p{{\tilde r}_1}}}{{{{\left( {\tilde r_1^2 + \tilde r_2^2} \right)}^2}}} + 2\left( {\tilde r_1^{} - {q_1}} \right) = 0,\\
				\dfrac{{\partial \tilde{f}\left( {{{\tilde r}_1},{{\tilde r}_2}} \right)}}{{\partial {{\tilde r}_2}}} = \dfrac{{ - 2p{{\tilde r}_2}}}{{{{\left( {\tilde r_1^2 + \tilde r_2^2} \right)}^2}}} + 2\left( {\tilde r_2^{} - {q_2}} \right) = 0.
	\end{align}
\end{subequations}
After some mathematical operations, we obtain ${q_2}\tilde r_1^{} = {q_1}\tilde r_2^{}$.
Now, consider following discussions:

(a) If $q_1=q_2=0$, according to \eqref{eq46}, we have $\tilde r_1^{2}+\tilde{r}_2^2=\sqrt{p}$. Without prior information, the optimal solution that satisfies $\tilde r_1^{2}+\tilde{r}_2^2=\sqrt{p}$ can be randomly generated.

(b) If $q_1\neq0,~q_2=0$, we know $\tilde r_2=0$. Substituting $\tilde r_2=0$ into \eqref{eq46} produces $\tilde r_1^4 - {q_1}\tilde r_1^3 - p = 0$ which is a quartic polynomial equation w.r.t. $\tilde{r}_1$. 
Thus, the solution can be easily obtained by the formula of computing roots \cite{BS3} or software like MATLAB.
Note that there may be more than one real root of the equation, and we need to select the one that minimizes $\tilde{f}(\tilde{r}_1, \tilde{r}_2) $.

(c) If $q_1=0,~q_2\neq0$, this case is similar to case (b). We know $\tilde r_1=0$ and need to solve $\tilde r_2^4 - {q_2}\tilde r_2^3 - p = 0$.

(d) If $q_1\neq0,~q_2\neq0$, this is the most normal case. According to \eqref{eq46}, the optimal $\tilde{r}_1$ can be obtained by solving
\begin{equation}
	\tilde r_1^4 - {q_1}\tilde r_1^3 - \frac{{pq_1^4}}{{{{\left( {q_1^2 + q_2^2} \right)}^2}}} = 0.
\end{equation}
Then, the  optimal $\tilde{r}_2$ is given by  $\tilde{r}_2 = q_2\tilde{r}_1/q_1$.

Finally, the $\mathbf{r}$ is updated as $\mathbf{r}= \mathbf{U}\mathbf{\tilde{r}}$.

\begin{algorithm}[t]
	
	\caption{{\color{black} Alternatin\emph{g} wavefo\emph{r}m and filt\emph{e}r d\emph{e}sign for QSINR maximiza\emph{t}ion (GREET)}} 
	\label{alg:2} 
	\begin{algorithmic}[1] 
		\Require  $N_t$, $N_r$, $L$, $\sigma^2$, $\{\theta_{0},\sigma_0^2,\alpha_0\}$, $\{\theta_k, \delta_k,\sigma_k^2\}_{k=1}^K$, $\{\rho_1,\rho_2\}$, $I_{\rm AltOpt}^{\max}$, $I_{\rm ADMM}^{\max}$.		
		\State $i \leftarrow0$.
		\State Randomly initialize $ \mathbf{s} \in \mathcal{X}^{N_tL \times 1} $.
		\State Randomly initialize $ \mathbf{t},\mathbf{r} \in \{ \frac{-1}{\sqrt{2N_tL}} , \frac{1}{\sqrt{2N_tL}} \}^{2N_tL \times 1} $.
		\State ${\bf{A}}\left( \theta_0  \right) \leftarrow {{\bf{I}}_L} \otimes {{\bf{a}}_r}\left( \theta_0  \right){\bf{a}}_t^T\left( \theta_0  \right)$.
		\Repeat
			\State $i \leftarrow i +1$.
			\State Compute $\bm{\Xi}(\mathbf{s})$ via \eqref{Xi25} or \eqref{Xi&Phi}.
			\State $	{\bf{w}} \leftarrow \frac{{{{\left( {{\bf{\Xi }}\left( {\bf{s}} \right) + {\bf{I}}_{N_rL}} \right)}^{ - 1}}{\bf{A}}\left( {{\theta _0}} \right){\bf{s}}}}{{{{\bf{s}}^H}{{\bf{A}}^H}\left( {{\theta _0}} \right){{\left( {{\bf{\Xi }}\left( {\bf{s}} \right) + {\bf{I}}_{N_rL}} \right)}^{ - 1}}{\bf{A}}\left( {{\theta _0}} \right){\bf{s}}}}$.
			\State $\mathbf{\Gamma}(\mathbf{w}) \leftarrow \mathbf{A}^H(\theta_0)\mathbf{w}\mathbf{w}^H\mathbf{A}(\theta_0)$.
			\State Compute $\bm{\Phi}(\mathbf{w})$ via \eqref{Phi27} or \eqref{Xi&Phi}.
			\State Transform $\mathbf{\Gamma}(\mathbf{w})$ and $\bm{\Phi}(\mathbf{w})$ into real-valued form.
			\State EVD  $\mathbf{\tilde{\Phi}}(\mathbf{w}) \leftarrow {\bf{{ P}\Pi }}{{\bf{P}}^T}$ and ${\bf{\tilde \Gamma }}\left( {\bf{w}} \right) \leftarrow {\bf{U\Lambda }}{{\bf{U}}^T}$.
			\State $j \leftarrow 0$, $\mathbf{u}_1\leftarrow \mathbf{0}_{2N_tL}$, and $\mathbf{u}_2\leftarrow \mathbf{0}_{2N_tL}$.
			\Repeat
				\State	$j \leftarrow j + 1$.
				\State Compute $\mathbf{\tilde{s}}$ by \eqref{update_s}.
				\State $\mathbf{g}\leftarrow \mathbf{P}^T (\mathbf{\tilde{s}} - \mathbf{u}_1) $.
				\State Compute $\nu$ by solving \eqref{BS1} via the BSM.
				\State Compute $\mathbf{\tilde{t}}$ by \eqref{eq37}, and then  $\mathbf{t} \leftarrow \mathbf{P} \mathbf{\tilde{t}}$.
				\State $\mathbf{q} \leftarrow \mathbf{U}^T(\mathbf{\tilde{s}} -\mathbf{u}_2 )$.
				\State $\tilde{r}_k\leftarrow q_k,~k=3,4...,2N_tL$.
				\State Compute $\tilde{r}_1, \tilde{r}_2$ by solving  \eqref{ADMM_r3}.
				\State $\mathbf{r} \leftarrow \mathbf{U}\mathbf{\tilde{r}}$.
				\State $\mathbf{u}_1 \leftarrow \mathbf{u}_1 + \mathbf{t}- \mathbf{\tilde{s}}$.
				\State $\mathbf{u}_2 \leftarrow \mathbf{u}_2 + \mathbf{r}- \mathbf{\tilde{s}}$.
			\Until $j=I_{\rm ADMM}^{\max}$, output $\mathbf{\tilde{s}}$, $\mathbf{t}$, and $\mathbf{r}$.
			\State Transform $\mathbf{\tilde{s}}$ into complex-valued form $\mathbf{s}$.
		\Until $i=I_{\rm AltOpt}^{\max}$.
		\Ensure	the transmit waveform $\mathbf{s}$ and the receive filter $\mathbf{w}$. 
	\end{algorithmic} 
\end{algorithm}

\subsection{Complexity and Convergence Analysis}
In \textbf{Algorithm 1}, we summarize the proposed algorithm termed GREET for solving the joint design problem \eqref{problem1}.
\subsubsection{Computational complexity}
As shown in \textbf{Algorithm 1}, the inner-loop is the implementation of the ADMM algorithm. 
{\color{black} The main computations include some matrix multiplications, such as $\mathbf{t}= \mathbf{P}\mathbf{\tilde{t}}$ and $\mathbf{r}= \mathbf{U}\mathbf{\tilde{r}}$, where each of them requires $(2N_tL)^2$ multiplications. 
Therefore, the computational complexity of the inner-loop is $\mathcal{O}(N_t^2L^2)$.}
Note that, for the existing SDR and BCD methods, their computational complexities of the inner-loop  are $\mathcal{O}(N_t^{3.5}L^{3.5}+\mathcal{K}N_t^2L^2)$ and $\mathcal{O}(N_t^{3}L^{3})$, respectively, where $\mathcal{K}$ is the number of randomizations for the SDR method.

For the outer-loop, {\color{black} i.e., iterating between $\mathbf{w}$ and $\mathbf{s}$}, the main computations include the matrix inversion and EVD operations. Thus, the computational complexity of the proposed GREET is $\mathcal{O}(   N_r^3 L^3+N_t^3 L^3 + I_{\rm ADMM}^{\max} N_t^2L^2)$, where $ I_{\rm ADMM}^{\max}$ is the allowed number of iterations in the ADMM method. 

\subsubsection{Convergence of the ADMM algorithm}
We use the following proposition to show the convergence of the ADMM algorithm.

\textbf{Proposition 2:} Let $\{\mathbf{\tilde{s}}^{(j+1)},\mathbf{t}^{(j+1)},\mathbf{r}^{(j+1)} \}$ be the resulting $\{\mathbf{\tilde{s},t,r} \}$ for the $(j+1)$-th iteration of the ADMM method, and define residual terms $\mathbf{d}^{(j+1)} \triangleq \mathbf{\tilde{s}}^{(j+1)}-\mathbf{\tilde{s}}^{(j)}$, $\mathbf{c}_1^{(j+1)}\triangleq \mathbf{t}^{(j+1)} - \mathbf{\tilde{s}}^{(j+1)}$, and 
 $\mathbf{c}_2^{(j+1)}\triangleq \mathbf{r}^{(j+1)} - \mathbf{\tilde{s}}^{(j+1)}$.
 Then $\mathbf{\tilde{s}}^{(j+1)}$ converges to a KKT point of the problem \eqref{subproblem_sr} when $\mathbf{d}^{(j+1)} \rightarrow \mathbf{0}_{2N_tL}$, $\mathbf{c}_1^{(j+1)} \rightarrow \mathbf{0}_{2N_tL}$, and $\mathbf{c}_2^{(j+1)} \rightarrow \mathbf{0}_{2N_tL}$.
\begin{IEEEproof}
	see \textbf{Appendix D}.
\end{IEEEproof}

\section{Simulation Results}
In this section, several sets of simulations are provided to illustrate the performance of the proposed scheme. 
\color{black}
Without loss of generality, in all simulations, the noise power is set as $\sigma^2 =1$, the target is assumed to be nonfluctuating, and the transmit and receive antenna arrays are uniform linear arrays with half-wavelength antenna spacing.
The  theoretical QSINR is computed by \eqref{eq18}, and the practical QSINR is obtained via estimating \eqref{QSINR_prac}, i.e.,
	\begin{equation}
	{\rm QSNR_{MC}} = \frac{1}{K_{{\rm MC}}} \sum\limits_{k=1}^{K_{{\rm MC}}} \frac{|\mathbf{w}^H \mathcal{Q}( \mathbf{h}_1 + \mathbf{v}_k) |^2}{ |\mathbf{w}^H \mathcal{Q}(\mathbf{h}_0 + \mathbf{v}_k) |^2  } -1,
	\end{equation} 
	where $K_{{\rm MC}}$ denotes the number of Monte Carlo tests, $\mathbf{v}_k$ is a sample of $\mathbf{v}$ for the $k$-th test.
	The number of Monte Carlo tests is $10^4$ unless otherwise specified.
	In the noise-only case, the transmit waveform is set according to \eqref{varphi_solution} and the receive filter is the matched filter.
	In the presence of interferences, the GREET framework is applied to jointly design $\mathbf{w}$ and $\mathbf{s}$.

 \color{black}
 
\subsection{Noise-Only Case}
In the first example, we consider the noise-only case, i.e., in the absence of interference. 
\color{black} 
We set the number of transmit antennas $N_t=8$ and the target angle $\theta_0=22^{\circ}$. The complex amplitude of the target is set as $\alpha_0 = 10$. 
We consider four different combinations of DACs/ADCs, i.e., 
\begin{itemize}
	\item[C1:] Infinite-bit ($\infty$-bit) DACs/ADCs;
	\item[C2:] One-bit DACs and $\infty$-bit ADCs;
	\item[C3:] $\infty$-bit DACs and one-bit ADCs;
	\item[C4:] One-bit DACs/ADCs.
\end{itemize}
 \color{black}

\begin{figure}[!t]
	\centerline{\includegraphics[width=0.4\textwidth]{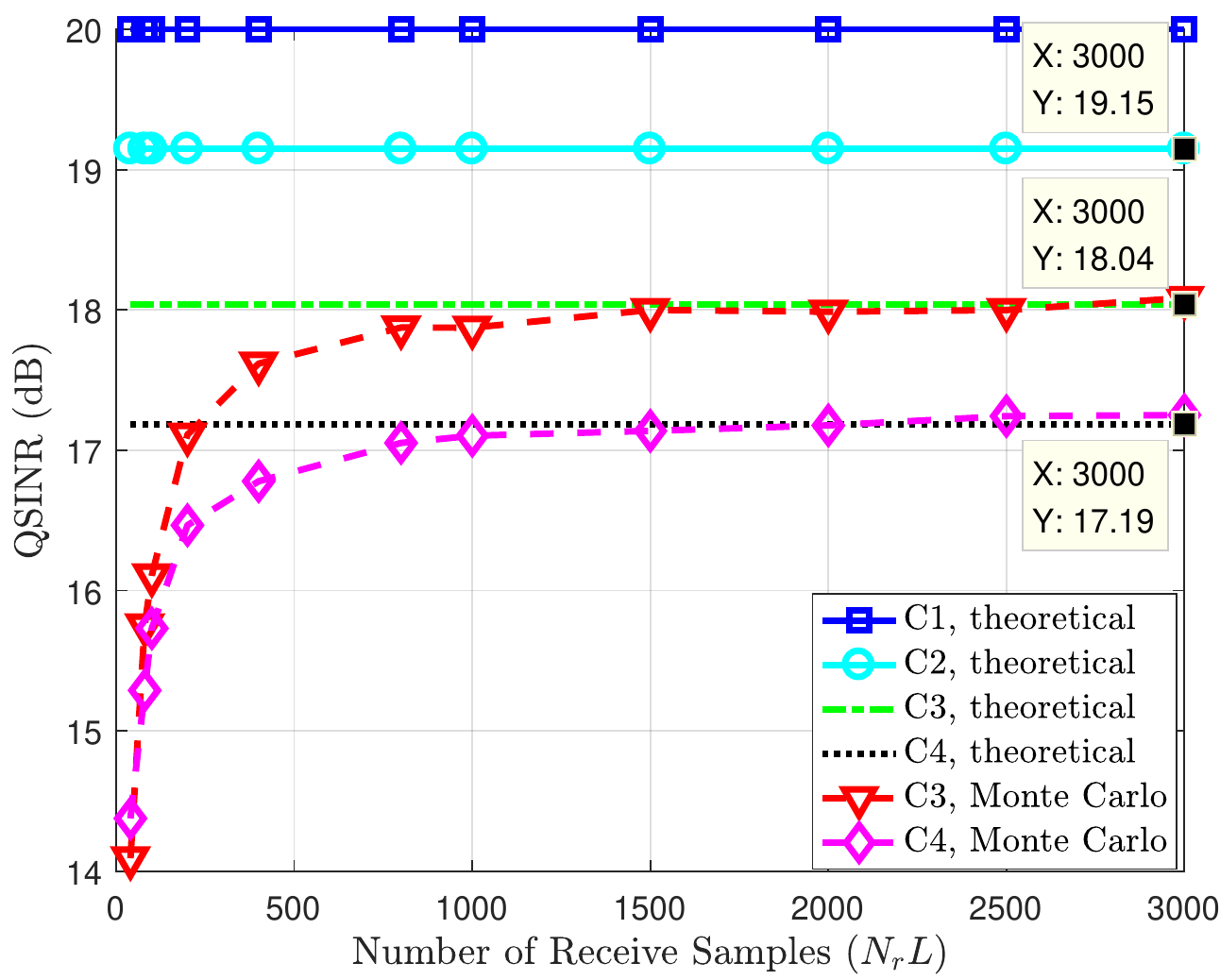}}
	\caption{The output QSINR (SNR) for different combinations of DACs/ADCs in the noise-only case, with $\theta_{0}=22^{\circ}$, $|\alpha_0|^2=20$dB, $N_t=8$, $N_r=5$, and $N_rL$ varies from 40 to 3000. }
	\label{fig1}
\end{figure}

\subsubsection{QSINR behavior}
In Fig. \ref{fig1}, we give the theoretical QSINR for C1-C4 according to the conclusion in Sec. IV-C, and in particular, for C3 and C4 which employ one-bit ADCs, the Monte Carlo tests are performed to estimate their QSINR.
As shown in Fig. \ref{fig1}, since the transmit waveform and the steering vectors are normalized, the maximum output QSINR (SNR) is $|\alpha_0|^2/\sigma^2=20$dB for the ideal case, i.e., $\infty$-bit DACs/ADCs.
In contrast to the ideal case, the losses of C2, C3, and C4 are $0.85$dB, $1.96$dB, and $2.81$dB, respectively. Moreover, we also notice that for a fixed $\alpha_0$,  as $N_rL$ increases, the gap between the theoretical QSINR and the one obtained by Monte Carlo tests becomes small.  
This is because the larger $N_rL$ results in a smaller input SNR per sample, i.e., ${\rm SNR}_{\rm per}=|\alpha_0|^2/N_rL$, leading to improved condition of the LIS assumption as well as the central-limit theorem. In fact, we find that when $N_rL\ge 1000$, i.e., ${\rm SNR}_{\rm per}\leq-10\text{dB}$, this gap is very small.

\begin{figure}[t]	
	
		\subfigure[]{
			\includegraphics[width=0.23\textwidth]{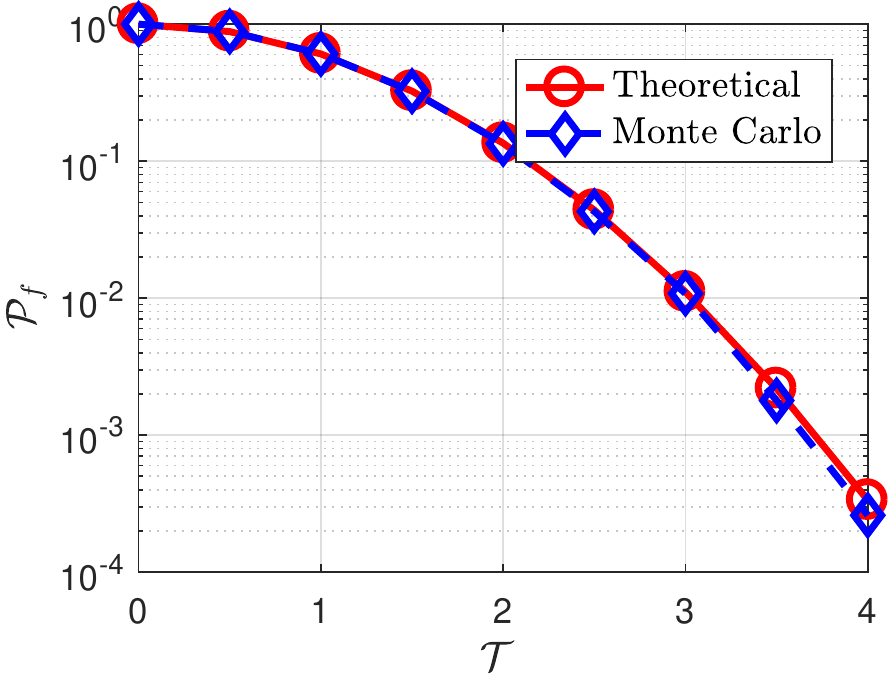} 
		}\subfigure[]{
			\includegraphics[width=0.23\textwidth]{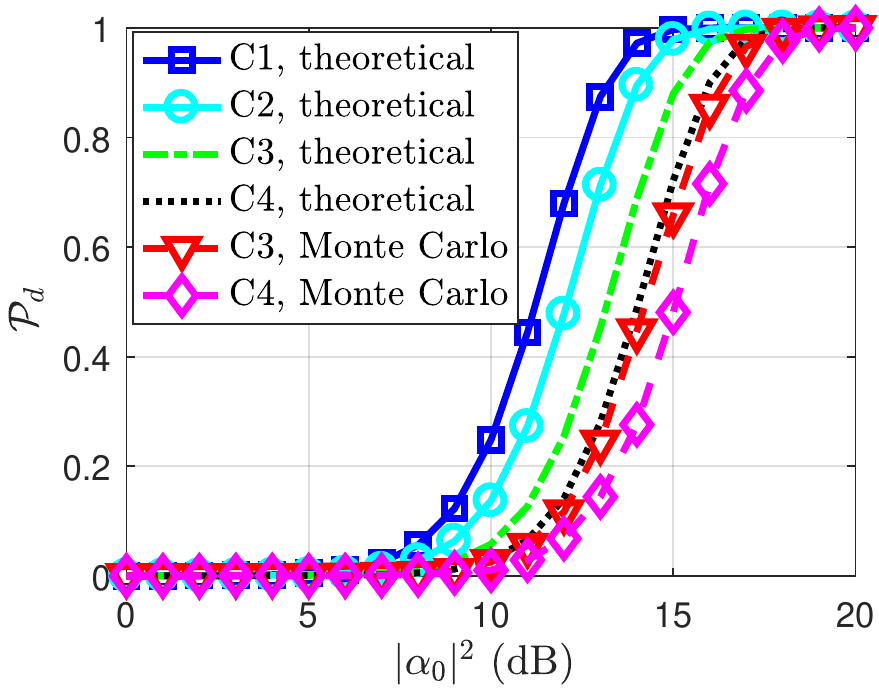} 
		}

		\subfigure[]{
			\includegraphics[width=0.23\textwidth]{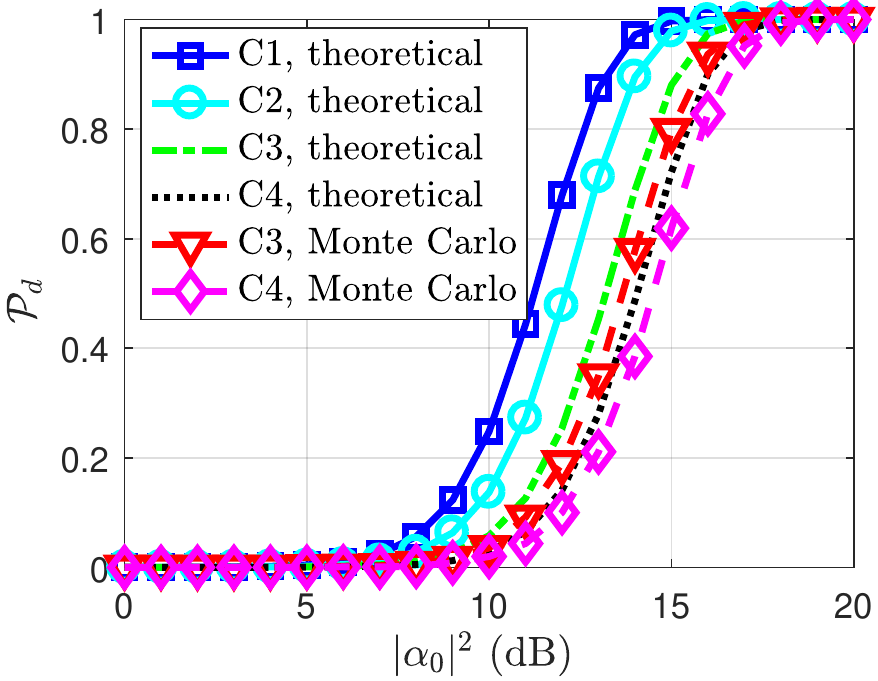} 
		}\subfigure[]{
			\includegraphics[width=0.23\textwidth]{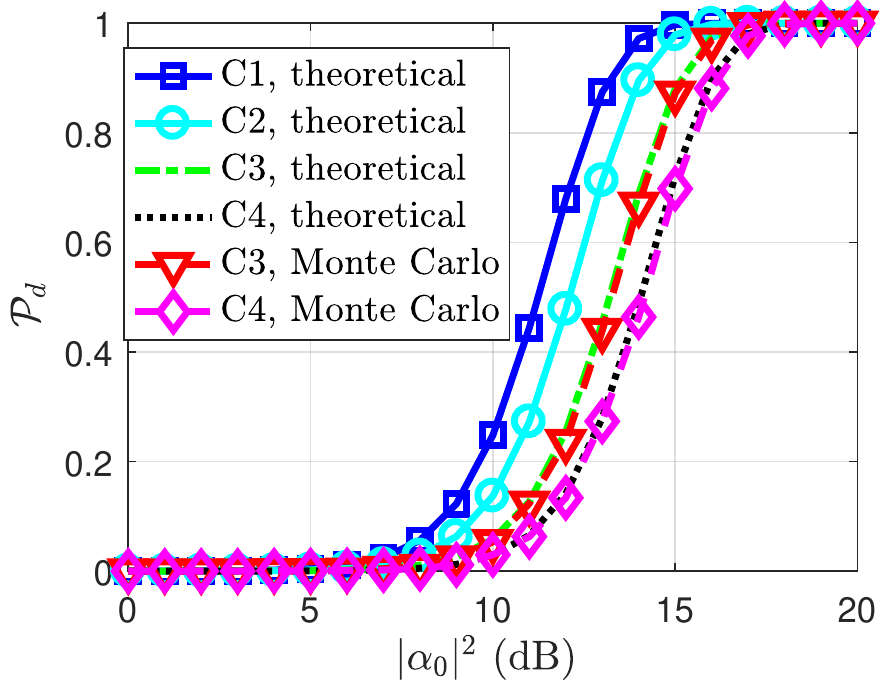} 
		}

	\caption{Detection performance in the noise-only case, with $\theta_{0}=22^{\circ}$, $N_t=8$ and $N_r=5$. (a) $\mathcal{P}_f$ versus $\mathcal{T}$, and the Monte Carlo curve is obtained by setting $L=100$ and performing $10^6$ independent tests; (b) $\mathcal{P}_d$ versus $|\alpha_0|^2$, $L=10$, $\mathcal{P}_f = 10^{-6}$; (c)  $\mathcal{P}_d$ versus $|\alpha_0|^2$, $L=20$, $\mathcal{P}_f = 10^{-6}$; (d)  $\mathcal{P}_d$ versus $|\alpha_0|^2$, $L=100$, $\mathcal{P}_f = 10^{-6}$. } %
	\label{fig2} 
\end{figure}

\subsubsection{Detection performance}
Next, Fig. \ref{fig2} shows the detection performance in the noise-only case, where (a) gives the probability of false-alarm versus the detection threshold $\mathcal{T}$, and (b)-(d) give the probability of detection versus the target power $|\alpha_0|^2$.
It can be observed from Fig. \ref{fig2}(a) that the Monte Carlo $\mathcal{P}_f$ is close to the theoretical one, which verifies the result in \eqref{P_f}.
Moreover, observing the results in Fig. \ref{fig2}(b)-(c), there exists performance difference between the theoretical and Monte Carlo $\mathcal{P}_d$.
This is because they employ a small number of receive samples, i.e., $N_rL=50$ and $N_rL=100$, which can not ensure a good condition for the LIS assumption and central-limit theorem.
In Fig. \ref{fig2}(d), as expected, the Monte Carlo $\mathcal{P}_d$ is close to its theoretical value when employing a larger $N_rL=500$.

\begin{figure}[!t]
	\centerline{\includegraphics[width=0.4\textwidth]{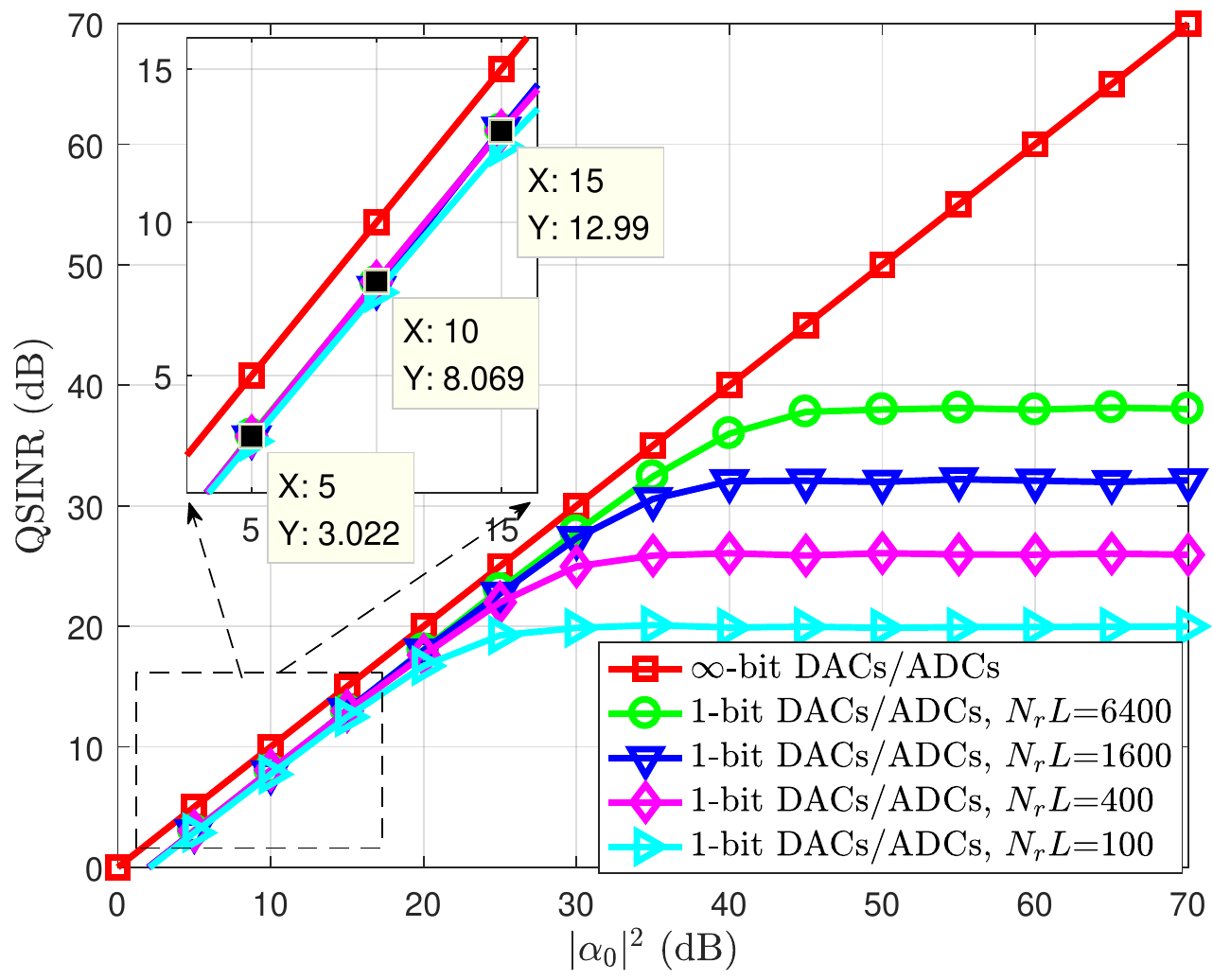}}
	\caption{The output QSINR (SNR) versus the target power $|\alpha_0|^2$ and the number of receive samples $N_rL$, where $N_t=8$ and $\theta_0=0^{\circ}$. }
	\label{fig3}
\end{figure}

\subsubsection{QSINR versus $|\alpha_0|^2$ and $N_rL$}
Finally, Fig. \ref{fig3} displays the QSINR (via Monte Carlo tests) versus the target power $|\alpha_0|^2$  and different $N_rL$.  
Since the transmit waveform and the steering vectors are normalized, the QSINR for $\infty$-bit converters is equal to $|\alpha_0|^2$.
In contrast to the ideal case, the QSINR loss in the low regime of $|\alpha_0|^2$ is close to the theoretical value, i.e., $1.96$dB (no transmit loss due to $\theta_{0}=0^{\circ}$).
However, in the high regime of $|\alpha_0|^2$, the QSINR loss is enlarged and and the QSINR approaches an asymptotic value.
Moreover, the larger $N_rL$,  the larger this asymptotic value is. 
Actually, this asymptotic value is $N_rL-1$, and the reason can be found from the definition of QSINR.
Specifically, the fraction in \eqref{QSINR_prac} is a general Rayleigh quotient $\frac{ \mathbf{w}^H \mathbf{R}_1  \mathbf{w} }{\mathbf{w}^H \mathbf{R}_0  \mathbf{w}} $, where $\mathbf{R}_1 = \mathbb{E}\{ \mathcal{Q}( \mathbf{h}_1 + \mathbf{v})  \mathcal{Q}^H( \mathbf{h}_1 + \mathbf{v})  \}$ and $\mathbf{R}_0 = \mathbb{E}\{ \mathcal{Q}( \mathbf{h}_0 + \mathbf{v})  \mathcal{Q}^H( \mathbf{h}_0 + \mathbf{v})  \}$. 
In the noise-only case, we have $\mathbf{R}_0=2\mathbf{I}_{N_rL}$, and  $\mathbf{R}_1$ is positive semidefinite with all diagonal elements being $2$. Thus, the maximum value of this Rayleigh quotient equals to $N_rL$.


\subsection{Comparison between GREET and Existing Methods}
\begin{figure}[!t]

		\subfigure[]{
			\includegraphics[width=0.23\textwidth]{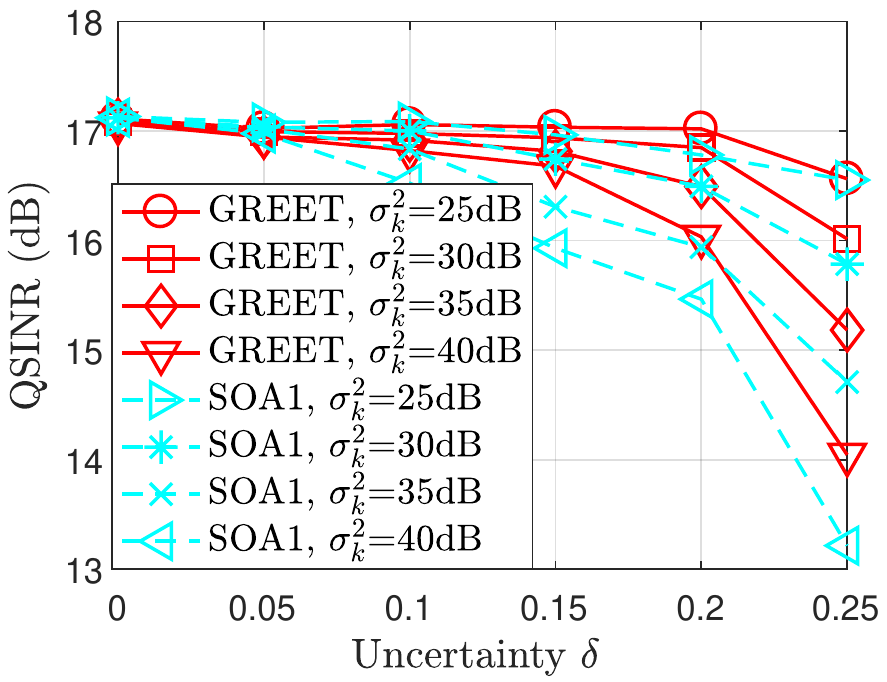} 
		}\subfigure[]{
			\includegraphics[width=0.23\textwidth]{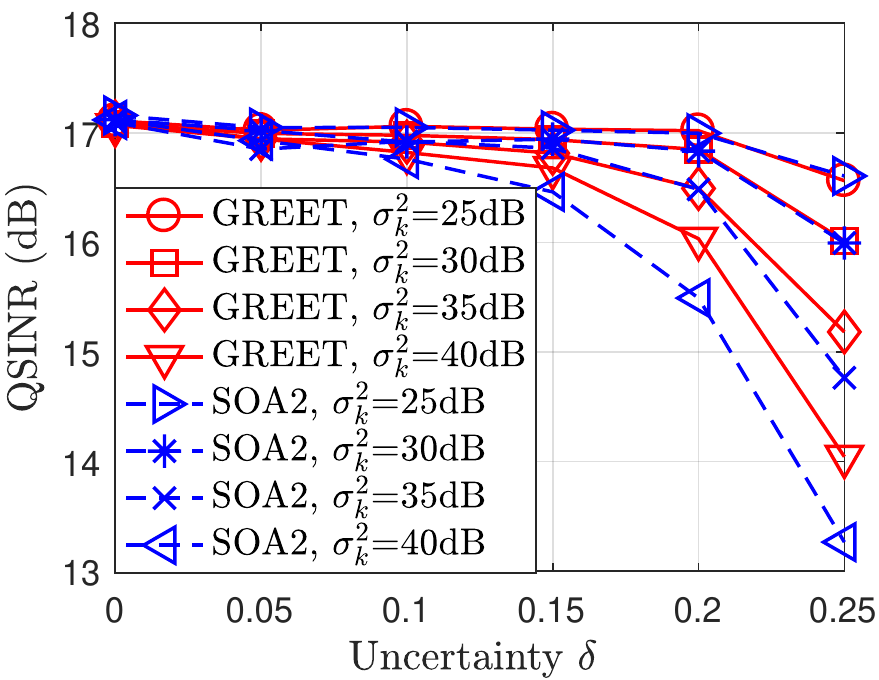} 
		}

		\subfigure[]{
			\includegraphics[width=0.23\textwidth]{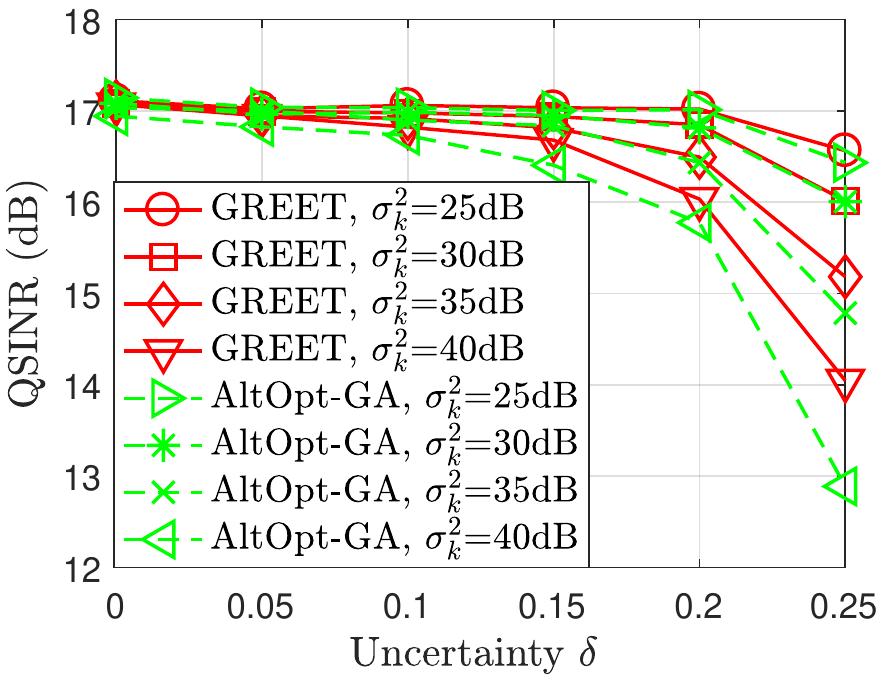} 
		}\subfigure[]{
			\includegraphics[width=0.23\textwidth]{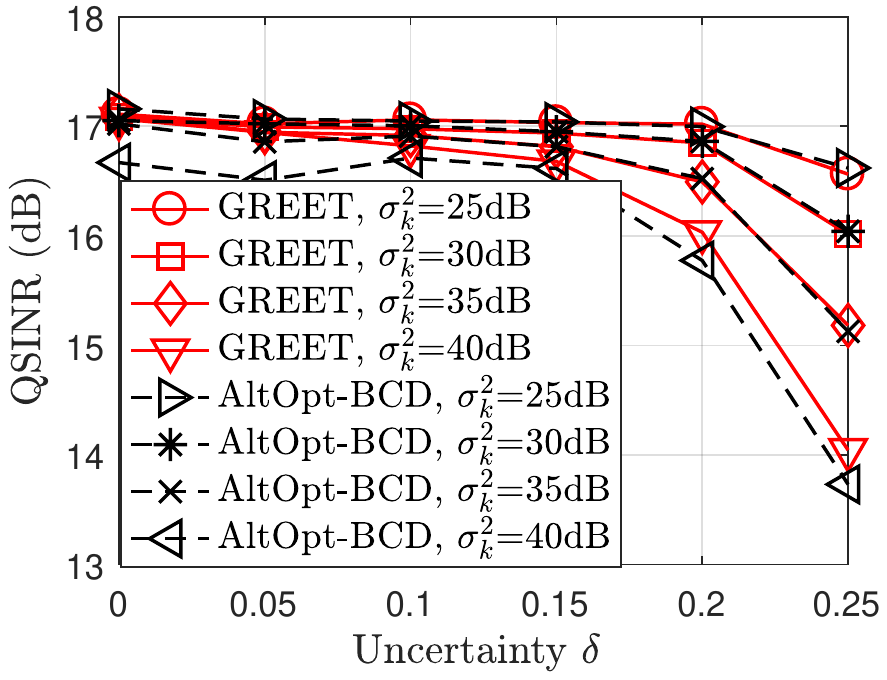} 
		}
	\caption{QSINR for one-bit MIMO radar in the presence of interferences, where $\{N_t,N_r,L\}=\{10,8,16\}$,  $\{ \theta_0, |\alpha_0|^2\}=\{34^{\circ},20 {\rm dB} \}$, two interferences $\{\overline{\omega}_1,\overline{\omega}_2  \}=\{\sin -50^{\circ},\sin 10^{\circ} \}$, $\sigma_k^2=25{\rm dB},30{\rm dB},35{\rm dB},40{\rm dB} (k=1,2)$, and $\delta_1=\delta_2=\delta \in [0,0.25]$. Comparison between GREET and (a)  SOA1; (b) SOA2; (c) AltOpt-GA; (d) AltOpt-BCD.}
	\label{fig4}
\end{figure}

The purpose of this example is to compare the difference between the proposed GREET and existing methods.
Specifically, we set $\{N_t,N_r,L\}=\{10,8,16\}$ and consider a target with $\{ \theta_0, |\alpha_0|^2\}=\{34^{\circ},20 {\rm dB} \}$.
Two interferences are considered, with their normalized directions $\{\overline{\omega}_1,\overline{\omega}_2  \}=\{\sin -50^{\circ},\sin 10^{\circ} \}$.
Moreover, we assume that the power and angle uncertainty of different interferences are equal, i.e., $\sigma_1^2=\sigma_2^2$ and $\delta_1=\delta_2=\delta$.
When implementing the GREET framework, the penalty variables of the ADMM method are $\rho_1=2$ and $\rho_2=30$.
The inner-loop is terminated after $I^{\max}_{\rm ADMM}=200$ iterations, and the outer-loop is terminated after $I^{\max}_{\rm AltOpt}=50$ iterations.
In addition, we also perform the following methods for comparison purpose:
\begin{itemize}
    \item SDR based sequential optimization algorithms \cite{B11}, labeled as ``SOA1'' and ``SOA2'', where the number of randomization trials for SOA1 is $20000$ and for SOA2 is $200$.
    \item Get the constant modulus (CM) solution of \eqref{subproblem_s} via the gradient approach (GA) \cite{6104176}, then convert it to one-bit waveform, labeled as ``AltOpt-GA''.
    \item Apply the block coordinate descend (BCD) method \cite{Add2} to solve \eqref{subproblem_s}, labeled as ``AltOpt-BCD ''.
\end{itemize}

\subsubsection{Comparison of QSINR}
Fig. \ref{fig4} shows the theoretical output QSINR obtained by different optimization methods. 
As expected, the values of the output QSINR show an decreasing trend as the power and angle uncertainty of interferences grow.
When $\delta$ and $\sigma_k^2$ $(k=1,2)$ are small, we find that the performance difference among these methods is insignificant.
Since all the involved methods employ the MVDR solution of $\mathbf{w}$, the reason might be that the degree-of-freedom of receive design is enough to ensure a good solution if both $\delta$ and $\sigma_k^2$ are small. 
The most interesting aspect of this graph is that, for large $\delta$ and $\sigma_k^2$, the proposed GREET outperforms the other alternative methods in terms of the output QSINR.

\renewcommand\arraystretch{1.2}
\begin{table}[htbp]
	\caption{The operation time for different optimization methods.}
	\label{table1}
	\begin{center}
		{
		\begin{tabular}{|c | c |c c c c |}
			\hline 	
			    & $L$  & 16 &32 &64 &128  	\\
			\hline			
			\multirow{5}{*}{time (s)}	&  SOA1 	    		& 90   & 4315 & - & -       	\\			
			\cline{2-6} 				&  SOA2  		        & 354  & 3791 & - & -  	\\
			\cline{2-6} 				&  AltOpt-GA  		    & 54.5 & 1537 & - & -  	\\
		    \cline{2-6} 				&  AltOpt-BCD  		    & 1.36 & 6.65 & 158 & 2140  	\\
			\cline{2-6} 				&  GREET  		        & 2.98 & 6.02 & 38.2 & 215 	\\			
			\hline
			\multicolumn{6}{|c|}{``-": non-execution due to forbidden computational cost} \\		
			\hline 
		\end{tabular}
		}
	\end{center}
\end{table}
\subsubsection{Comparison of computational efficiency}
Next, table \ref{table1} analyzes the performance of the GREET and other alternative methods in terms of the computational efficiency, where we employ the same simulation parameters as those of the last example except  $\sigma_1^2 = \sigma_2^2 = 30$dB, $\delta_1=\delta_2=0$, and $L=\{16,32,64,128\}$.
Each method involved is performed on a standard PC with 16GB RAM and 3.1 GHz CPU, and terminated after $50$ outer-loop iterations (iterating between $\mathbf{w}$ and $\mathbf{s}$). 
Due to the high computational complexity, the results show that the SDR based methods (i.e., SOA1 and SOA2), are computationally inefficient, especially for large $L$. 
Particularly, we find that the AltOpt-BCD method is more efficient than the GREET when $L=16$.
Nevertheless, as expected, since the ADMM has lower computational complexity than the BCD method, the GREET outperforms the AltOpt-BCD in terms of the computational efficiency when designing a long sequence.
Overall, from the results reported in Fig. \ref{fig4} and Tab. \ref{table1}, the superiority of the proposed GREET is demonstrated.

\begin{figure}[!t]
	{\centerline{\includegraphics[width=0.4\textwidth]{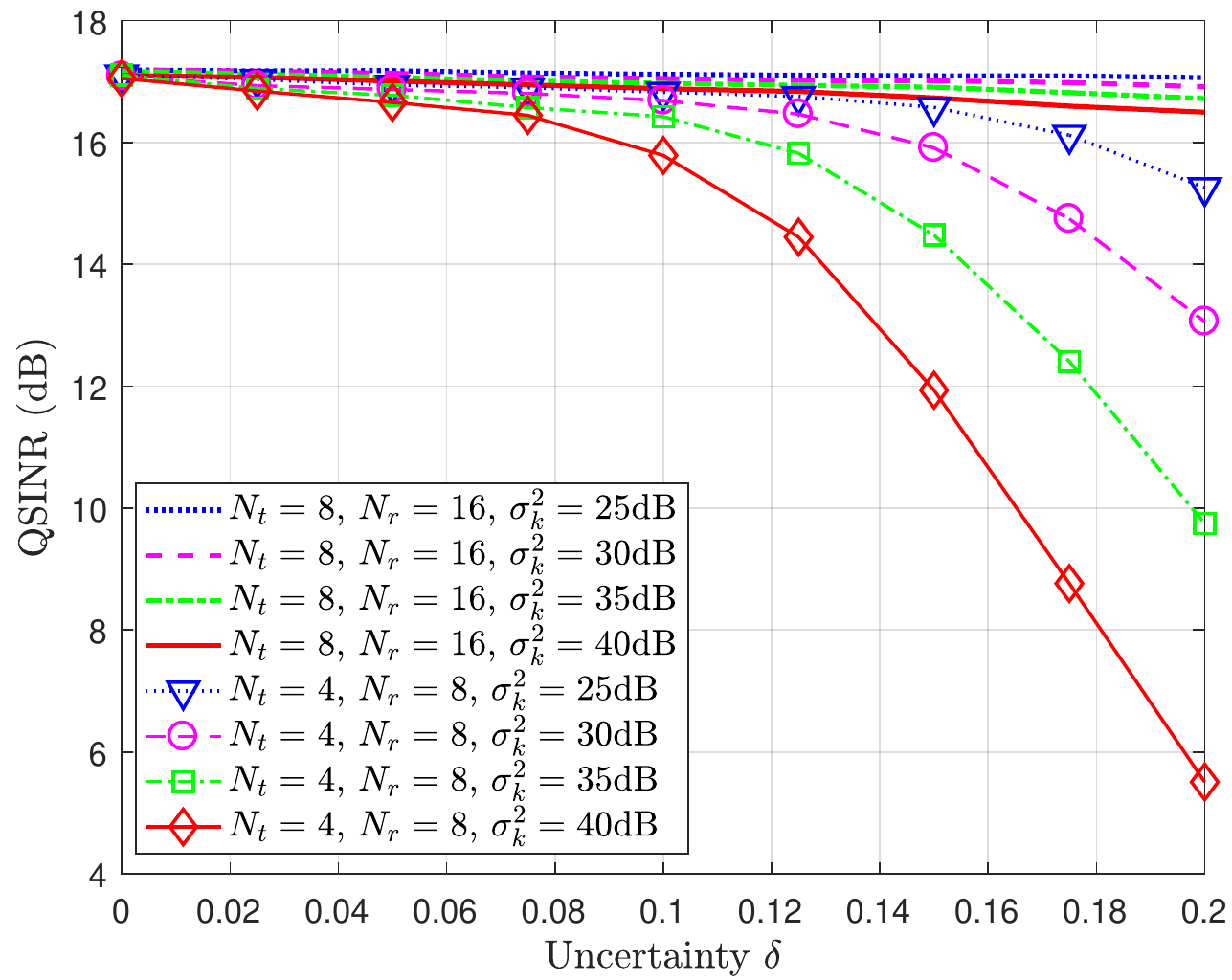}}}
	\caption{QSINR for one-bit MIMO radar in the presence of interferences, with $\{N_t,N_r\} = \{4,8\}$ or $\{8,16\}$, $L=16$, $\theta_0=-21^{\circ}$, $|\alpha_0|^2=20$dB, $\{\overline{\omega}_1,\overline{\omega}_2,\overline{\omega}_3\} = \{\sin -48^{\circ},\sin 5^{\circ}, \sin 54^{\circ}  \}$, $\sigma_1^2=\sigma_2^2=\sigma_3^2=25,30,35,40$dB, and $\delta_1=\delta_2=\delta_3=\delta\in [0,0.2]$.} 
	\label{fig6}
\end{figure}

\begin{figure}[!t]

		\subfigure[]{
			\includegraphics[width=0.24\textwidth]{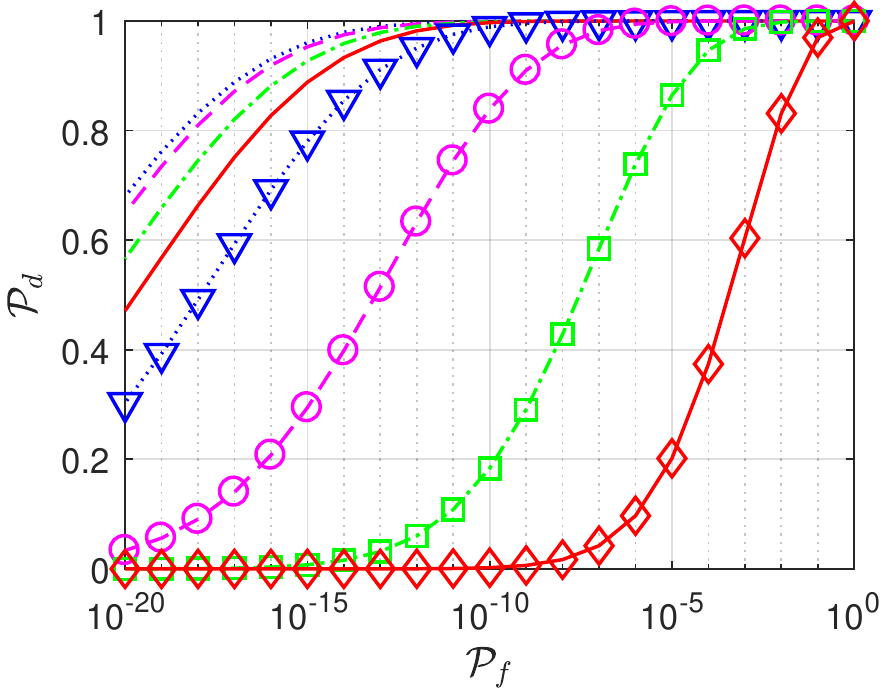} 
		}\subfigure[]{
			\includegraphics[width=0.24\textwidth]{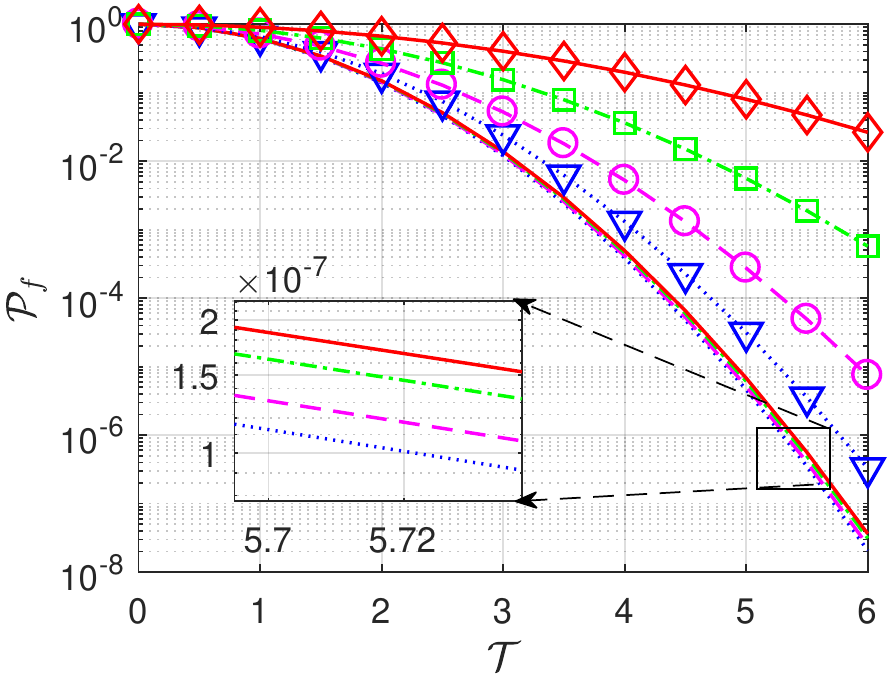} 
		}
\centering
		\subfigure[]{
			\includegraphics[width=0.4\textwidth]{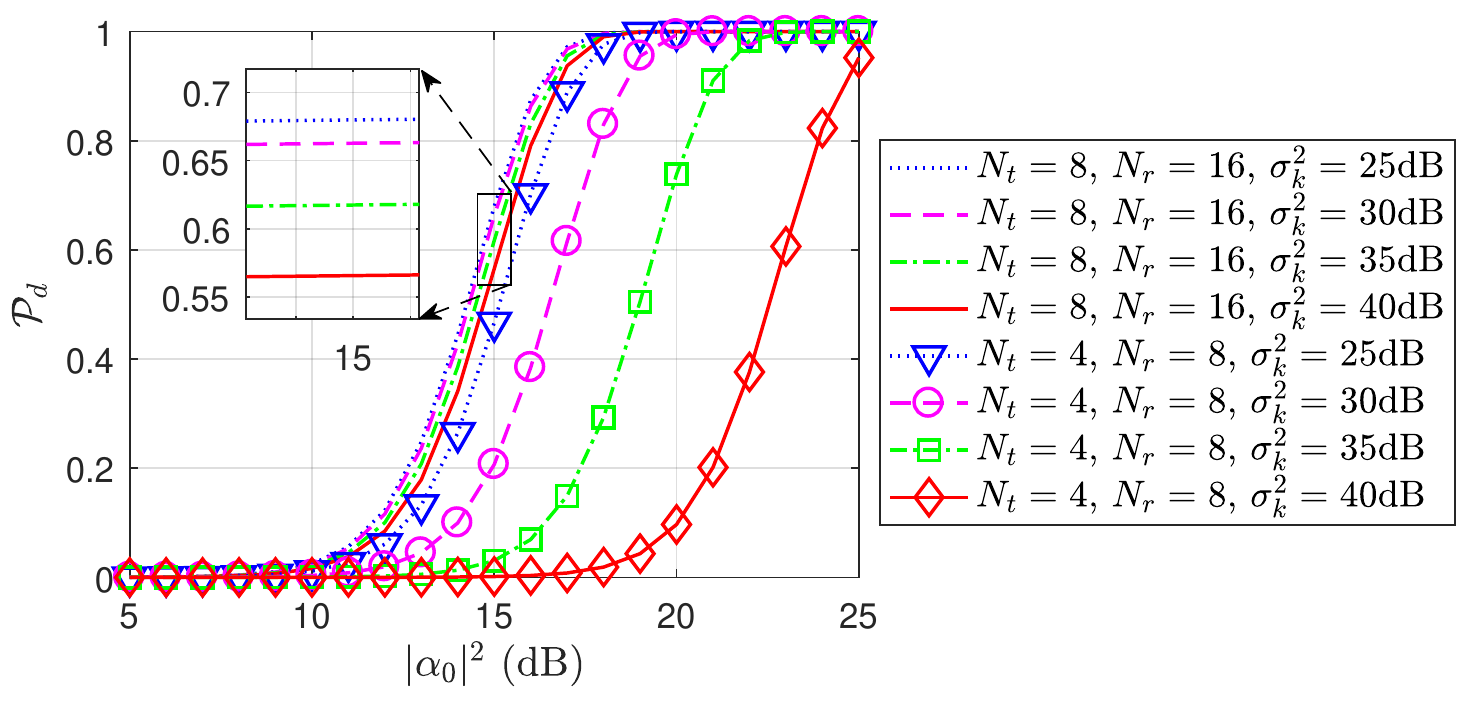} 
		}
	\caption{Detection performance for one-bit MIMO radar in Fig. \ref{fig6}, considering $\delta = 0.175$. (a) Receiver operating characteristic (ROC) curve, $|\alpha_0|^2=20$dB; (b) $\mathcal{P}_f$ versus $\mathcal{T}$, using normalized receive filter $\frac{\mathbf{w}}{||\mathbf{w}||_2}$; (c) $\mathcal{P}_d$ versus $|\alpha_0|^2$, $\mathcal{P}_f=10^{-6}$. }
	\label{fig7}
\end{figure}

\subsection{Detection Performance in the Presence of Interference}
In this example, we will evaluate the detection performance of one-bit MIMO radar.
Specifically, the target angle is $\theta_{0}=-21^{\circ}$.
We consider three interferences, with the average normalized angles $\{\overline{\omega}_1,\overline{\omega}_2,\overline{\omega}_3\} = \{\sin -48^{\circ},\sin 5^{\circ}, \sin 54^{\circ}  \}$, angle uncertainties $\delta_1=\delta_2=\delta_3=\delta\in[0,0.2]$, and powers $\sigma_1^2=\sigma_2^2=\sigma_3^2=25,30,35,40$dB.
When implementing the GREET, we set the penalty variables as $\{\rho_1,\rho_2\}=\{2,200\}$.
Moreover, the maximum iteration numbers of the inner-loop and outer-loop are $I^{\max}_{\rm ADMM}=200$ and $I^{\max}_{\rm AltOpt}=50$, respectively.

\subsubsection{QSINR behavior}
Using $\{\mathbf{w,s}\}$ designed by the GREET, Fig. \ref{fig6} shows the output QSINR versus the interference uncertainties for a target with $|\alpha_0|^2=20$dB.
It can be found that the output QSINR decreases with the increasing angle uncertainties and interference powers.
For $\{N_t,N_r\}= \{4,8\}$, one can observe that there is a remarkable performance loss if both the angle uncertainties and interference powers are large.
To enhance the interference suppression performance, we can increase the number of transmit and receive antennas.
As expected, when $\{N_t,N_r\}= \{8,16\}$, it can be seen that the interference suppression performance is significantly improved, especially for large $\delta$.

\subsubsection{Detection performance}
Next, considering $\delta=0.175$, Fig. \ref{fig7} illustrates the detection performance (theoretical) of one-bit MIMO radar in Fig. \ref{fig6}, where (a) gives the receiver operating characteristic (ROC) curve for $|\alpha_0|^2=20$dB, (b) presents the relation between the probability of false-alarm $\mathcal{P}_f$ and detection threshold $\mathcal{T}$, and (c) shows the probability of detection $\mathcal{P}_d$ versus the target power $|\alpha_0|^2$ with a fixed $\mathcal{P}_f=10^{-6}$.
Interestingly, one can observe that the detection performance is positively correlated with the QSINR.
This is in accordance with our analyses in Sec. III-A.

\begin{figure}[!t]	
		\centering	
		\subfigure[]{
			\includegraphics[width=0.24\textwidth]{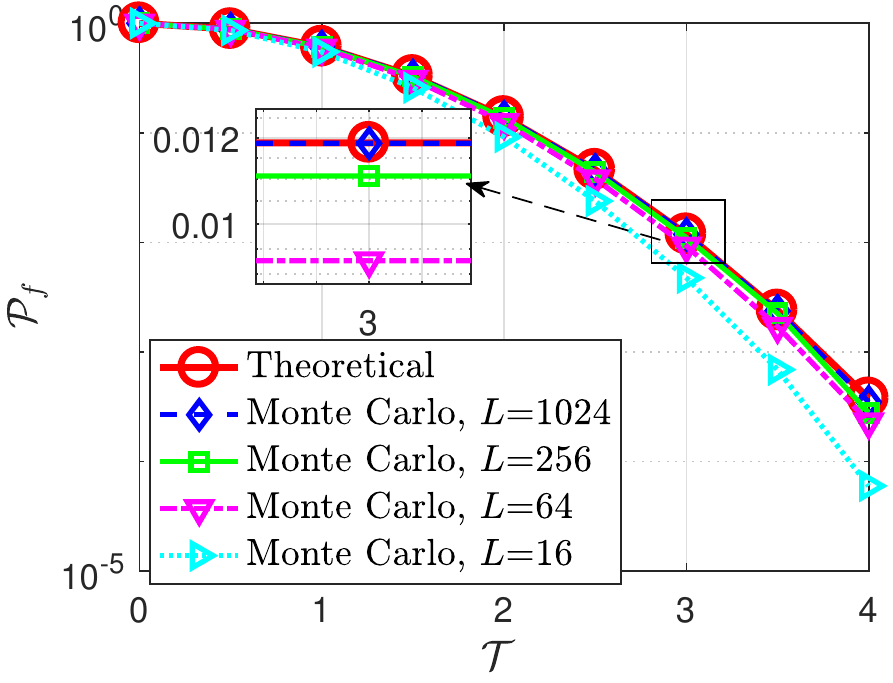} 
		}\subfigure[]{
			\includegraphics[width=0.24\textwidth]{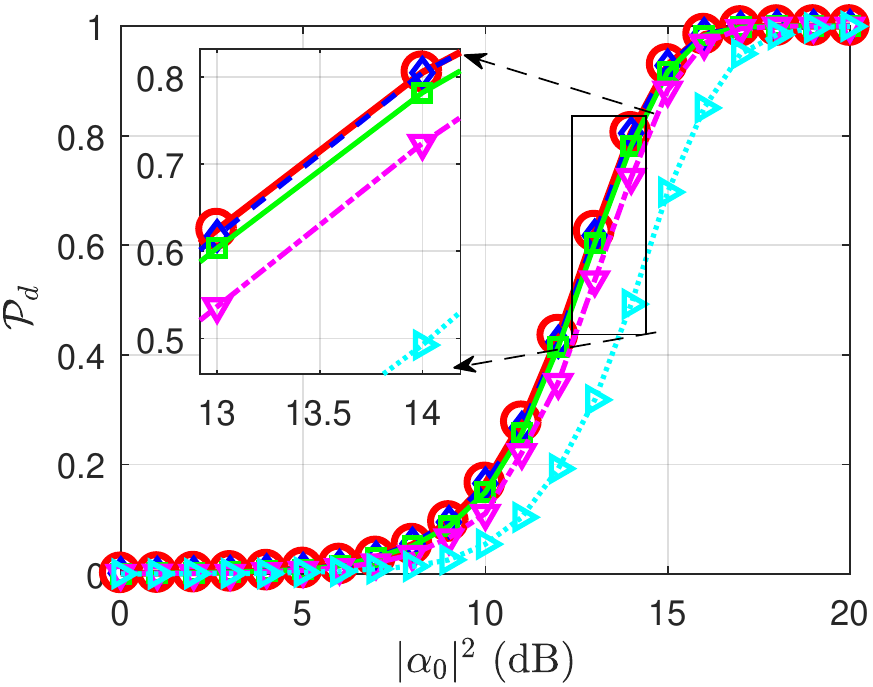} 
		}
	\caption{One-bit MIMO radar detection performance comparison between theoretical results and Monte Carlo results. (a) $\mathcal{P}_f$ versus $\mathcal{T}$, obtained after $10^{5}$ independent Monte Carlo tests; (b) $\mathcal{P}_d$ versus $|\alpha_0|^2$, $\mathcal{P}_f= 10^{-4}$. } %
	\label{fig7-1}
\end{figure}

Furthermore, we can perform Monte Carlo tests to evaluate the detection performance of one-bit MIMO radar in Fig. \ref{fig6}.
Specifically, we use the designed $\{\mathbf{w},\mathbf{s} \}$ with $\{N_t,N_r\}= \{8,16\}$, $\delta=0.1$, and $\sigma_k^2=30$dB ($k=1,2,3$).
In particular, we use different $L$, including $16,64,256,1024$, to represent the degree to which the receive signal satisfies the LIS assumption\footnote{Due to $\mathbf{s}^H\mathbf{s}=1$, increasing $L$ will decrease the power of a single sample of $\mathbf{s}$. Therefore, the larger $L$ indicates the better LIS assumption.}. 

We compare the theoretical and Monte Carlo detection performance for one-bit MIMO radar in Fig. \ref{fig7-1}.
Specifically, the left of Fig. \ref{fig7-1} presents the relation between the probability of false-alarm $\mathcal{P}_f$ and detection threshold $\mathcal{T}$, and the right one shows the probability of detection $\mathcal{P}_d$ versus the target power $|\alpha_0|^2$ at a fixed false-alarm $\mathcal{P}_f=10^{-4}$.
For a small $L$, e.g. $L=16$, we notice that there is a obvious difference between the theoretical and Monte Carlo curves.
This is because a small $N_rL$ can not provide with the LIS assumption a good condition.
Since the larger $L$ indicates the better LIS assumption, as can be seen, the gap between the theoretical and Monte Carlo results becomes small with $L$ increasing.

\subsection{Convergence Analysis}
\begin{figure}[!t]	
	
		\begin{minipage}[t]{1\linewidth}
		\centering	
		\subfigure[]{
			\includegraphics[width=0.8\textwidth]{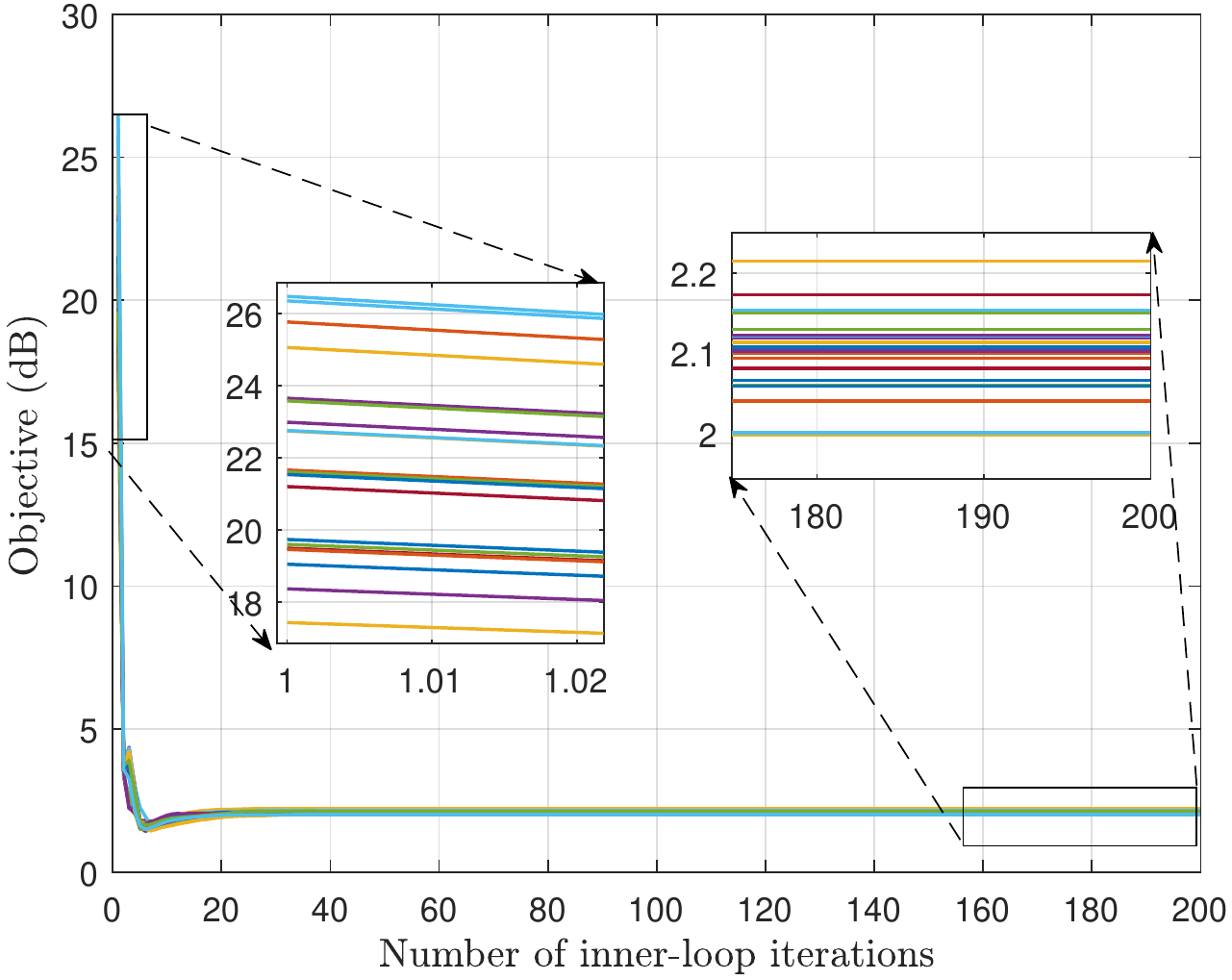} 
		}
		\end{minipage}
	\begin{minipage}[t]{1\linewidth}
		\centering	
		\subfigure[]{
			\includegraphics[width=0.45\textwidth]{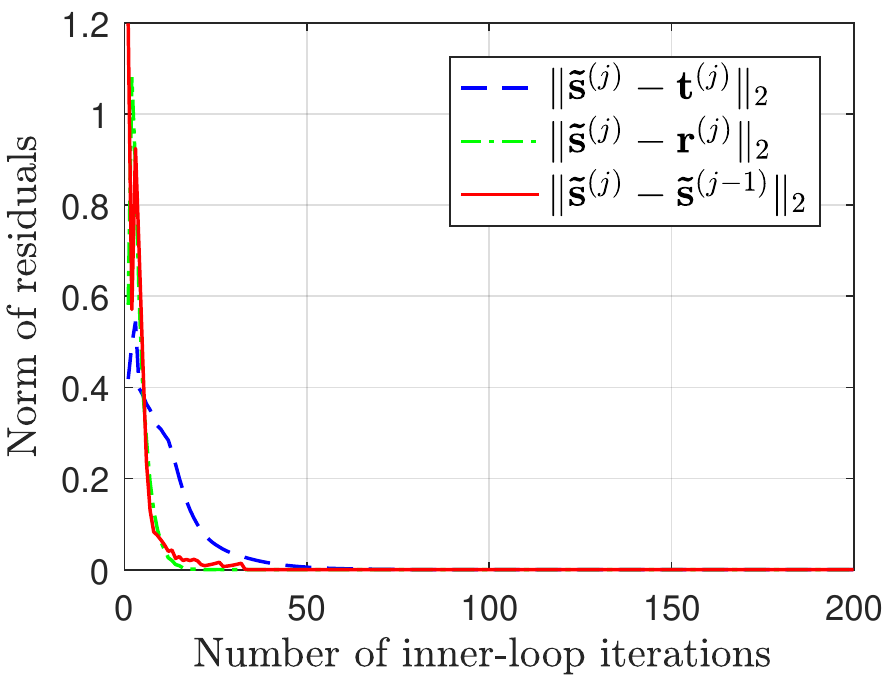} 
		}
		\subfigure[]{
			\includegraphics[width=0.45\textwidth]{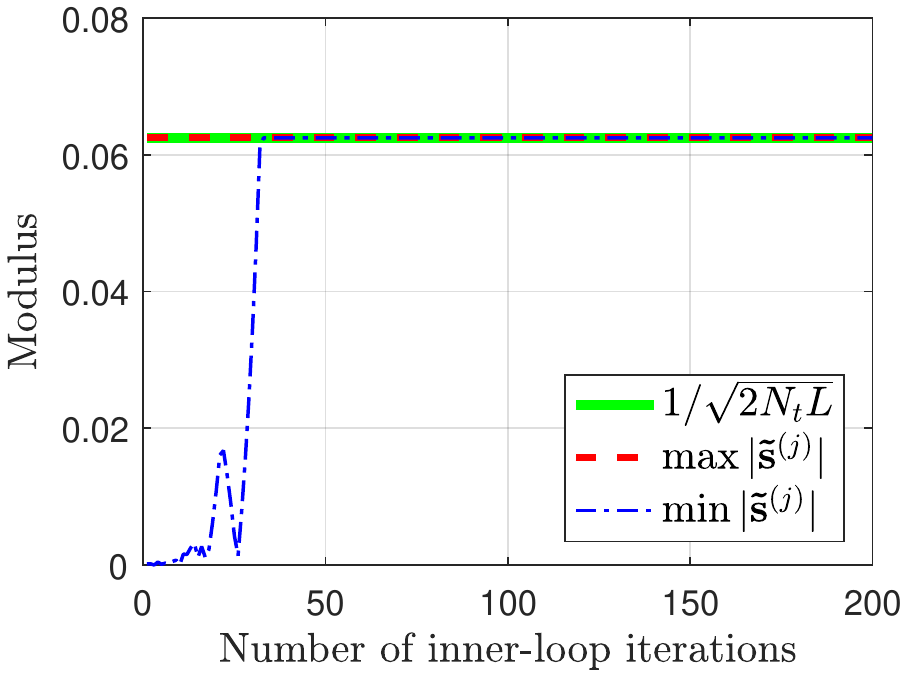} 
		}
	\end{minipage}
	\caption{Convergence property of the ADMM algorithm. (a) The objective value in problem \eqref{subproblem_sr} versus the iteration number with different initial points; (b) the norm of the residuals versus the iteration number; (c) the maximum and minimum modulus of $\mathbf{\tilde{s}}$ versus the iteration number. } %
\label{fig8}
\end{figure}

\begin{figure}[!t]
	\centerline{\includegraphics[width=0.4\textwidth]{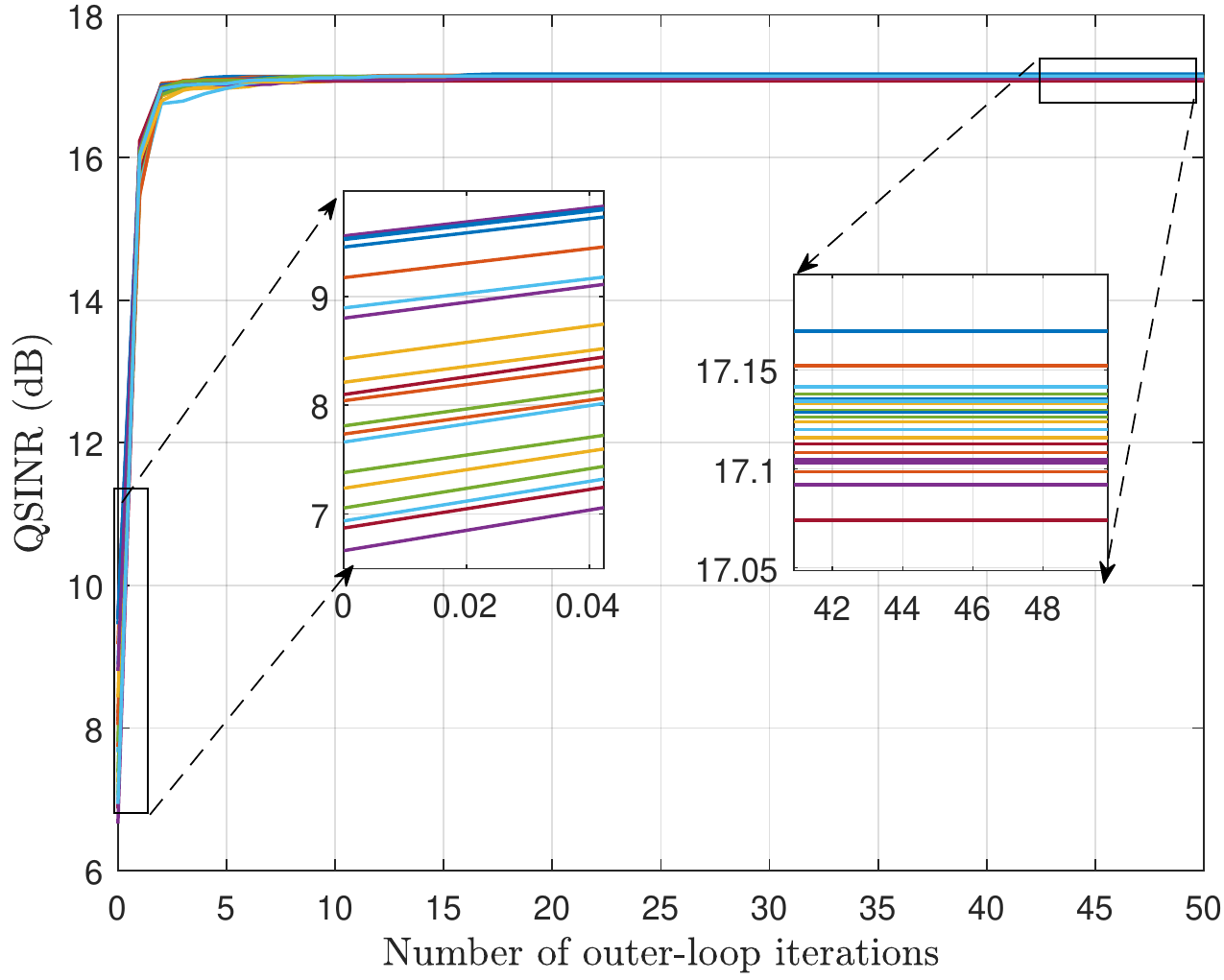}}
	\caption{The influence of different initial points on the outer-loop convergence.}
\label{fig9}
\end{figure}

Consider one of the simulation settings in Fig. \ref{fig6}, where $\{N_t,N_r,L \} =\{8,16,16\}$,  $\{\theta_0, |\alpha_0|^2 \} =\{-21^{\circ},20{\rm dB} \}$, $\{\theta_1,\theta_2,\theta_3 \} = \{-48^{\circ},5^{\circ},54^{\circ} \}$, $\delta =0$, $\sigma^2_k=30$dB ($k=1,2,3$), $\{\rho_1,\rho_2\}=\{2,200\}$, and $\{I^{\max}_{\rm ADMM},I^{\max}_{\rm AltOpt}\}=\{200,50 \}$.
In this example, we evaluate the convergence property of the proposed optimization method.

Firstly, we consider the inner-loop iteration, i.e., the ADMM method.
Specifically, given a fixed $\mathbf{w}$, we independently perform the ADMM method for $20$ times, where for each test we randomly set the initial $ \mathbf{t},\mathbf{r} \in \{ \frac{-1}{\sqrt{2N_tL}} , \frac{1}{\sqrt{2N_tL}} \}^{2N_tL \times 1} $. 
Fig. \ref{fig8}(a) presents the objective \eqref{subproblem_sr} versus the number of inner-loop iteration.
The result suggests that the different initial points can produce similar values of the objective.
Moreover, for one of these tests, we show the norms of iteration residuals and the minimum and maximum modulus of $\mathbf{\tilde{s}}^{(j)}$ versus the number of iterations ($j$) in Fig. \ref{fig8}(b) and Fig. \ref{fig8}(c).
As can be seen, the iteration residuals tend to be zero with the iteration number increasing. 
Based on the result in {\textbf{Proposition 2}}, we can infer that the obtained $\mathbf{\tilde{s}}^{(j)}$ converges to a KKT point of the problem \eqref{subproblem_sr}. Finally, the result in Fig. \ref{fig8}(c) shows that the modulus of $\mathbf{\tilde{s}}^{(j)}$ is close to a constant, i.e., $1/\sqrt{2N_tL}$. Thus, the waveform achieved by the proposed ADMM satisfies the constant modulus requirement.

Finally, we evaluate the convergence property of the outer-loop, i.e., iterating between $\mathbf{w}$ and $\mathbf{s}$.
Specifically, we independently perform the GREET for $20$ times.
For each test, Fig. \ref{fig9} presents the output QSINR versus the number of outer-loop iterations.
The result suggests that the different initial points can obtain a similar output QSINR.

\section{Conclusion}
{\color{black} In this paper, we have investigated the problem of detection performance analysis and joint design for one-bit MIMO radar.
Under the LIS assumption, the detection performance of one-bit MIMO radar in the presence of signal-dependent interferences has been mathematically derived.
Furthermore, we show that the detection performance is positively correlated with the QSINR, which is a function of the transmit waveform and receive filter. 
To achieve the best detection performance for the one-bit MIMO radar, a joint design problem has been formulated via maximizing the QSINR subject to a binary waveform constraint. This problem is then tackled by an AltOpt based method, termed as GREET, in which the receive filter and transmit waveform are obtained via a MVDR solution and the ADMM method, respectively.
In addition, the performance loss of one-bit MIMO radar has been evaluated, where for one-bit ADCs, the QSINR loss is about $1.96$dB in the low input SNR/INR regime, and for one-bit DACs, the worst QSINR loss in the noise-only case is less than $3$dB if the transmit waveform satisfies \eqref{varphi_solution}.
Finally, several representative simulations, involving both theoretical and Monte Carlo results, support our analyses and demonstrate the efficacy of the proposed method.

Nevertheless, this work only investigates the one-bit ADC/DAC MIMO radar for the target detection performance, more considerations of  the proposed one-bit MIMO radar for the estimation performance are still open problems. We shall pursue this  direction  in our future research work.}

\appendices
\color{black}
\section{Proof of Proposition 1}
Since ${x} \sim {\cal C}{\cal N}\left( {h,\sigma _{}^2} \right)$, we have $\Re \{ x \} \sim \mathcal{N}( \Re \{ h \}, 0.5 \sigma^2 )$ and	$\Im \{ x \} \sim \mathcal{N}( \Im \{ h \}, 0.5 \sigma^2 )$, respectively.
 
The probability of $\Re\{ y \} =1$ can be denoted as
\begin{equation}
\begin{aligned}
\mathcal{P}(\Re\{ y \} =1  ) &= \mathcal{P}(\Re \{ x \}>0  ) \\
&= \frac{1}{\sqrt{\pi \sigma^2}}  \int_{0}^{\infty} e^{-\frac{(t- \Re \{ h \})^2}{\sigma^2}} \text{d}t \\
&= \frac{1}{2} +  \frac{1}{2}{\rm erf} \left( \frac{\Re \{ h \}}{\sigma} \right).
\end{aligned}
\end{equation}
Similarly, the probability of $\Im\{ y \}=1$ is given by
\begin{equation}
\mathcal{P}(\Im\{ y \}=1) = \frac{1}{2} +  \frac{1}{2}{\rm erf} \left( \frac{\Im \{ h \}}{\sigma} \right).
\end{equation}
Moreover, the probabilities of  $\Re\{ y \} =-1$ and  $\Im\{ y \} =-1$ can be denoted as
\begin{subequations}
	\begin{align}
	&	\mathcal{P}(\Re\{ y \} =-1  ) = 1- \mathcal{P}(\Re\{ y \} =1  ), \\
	&	\mathcal{P}(\Im\{ y \} =-1  ) = 1- \mathcal{P}(\Im\{ y \} =1  ).
	\end{align}
\end{subequations}
Hence, the mean and variance of $y$ are given by
\begin{subequations}
\begin{align}
	\mathbb{E}\{ y \} &= \mathbb{E}\{ \Re \{ y\} \} + \jmath \mathbb{E}\{ \Im \{ y\} \} \\
&=  {\rm erf} \left( \frac{\Re \{ h \}}{\sigma} \right) +\jmath {\rm erf} \left( \frac{\Im \{ h \}}{\sigma} \right), \\
    \mathbb{D}\{ y \} &= \mathbb{E}\{ |y|^2 \}- |\mathbb{E}\{ y \}|^2=2- |\mathbb{E}\{ y \}|^2.
\end{align}
\end{subequations}

Next, since the function ${\rm erf}(x)$ is continuous and differentiable w.r.t. $x$, its  first-order Taylor expansion at $x=0$ is denoted as
\begin{equation}
\begin{aligned}
\label{taylor}
{\rm erf}(x)= {\rm erf}(0)+{\rm erf}'(0)x+o(x) 
=\frac{2}{\sqrt{\pi}}x+o(x),
\end{aligned}
\end{equation}
where ${\rm erf}'(x)=\frac{2}{\sqrt{\pi}}e^{-x^2}$ is the first-order derivative. 

Using the first-order approximation, we know
\begin{equation}
\begin{aligned}
	\mathbb{E}	\left\{ y \right\} &= \sqrt{\frac{4}{\pi \sigma^2}} (\Re \{ h \} + \jmath \Im \{ h \}) + o(\frac{\Re \{ h \}}{ \sigma }) + \jmath o(\frac{\Im \{ h \}}{ \sigma }) \\
	&= \sqrt{ \frac{4}{\pi\sigma^2} } h + o(\frac{h}{ \sigma }) = \sqrt{ \frac{4}{\pi\sigma^2} } h + o(\epsilon).
\end{aligned}
\end{equation}
Moreover, note that the approximation error of $	\mathbb{D}	\left\{ y \right\} $ is $|	\mathbb{E}	\left\{ y \right\} |^2$ satisfying
\begin{equation}
\begin{aligned}
\lim\limits_{ |\epsilon| \rightarrow 0 } \frac{ |	\mathbb{E}	\left\{ y \right\}|^2  }{ |\epsilon| } &= \lim\limits_{ |\epsilon| \rightarrow 0 }  \frac{ \frac{4}{\pi} |\epsilon|^2 + \sqrt{\frac{4}{\pi}} \Re \{ \epsilon o(\epsilon) \}  +|o(\epsilon)|^2   }{ |\epsilon| } \\
& =0.
\end{aligned}
\end{equation}
Therefore the approximation error of $	\mathbb{D}	\left\{ y \right\} $ is also $o(\epsilon)$, and the proof is completed.

\section{Derivation of $p_z(z|\mathcal{H}_0)$}
To simplify notation, we let $\varrho= 2\mathbf{w}^H\mathbf{w}$, $\bm{\Psi}= \bm{\beta \beta}^H/\varrho + \bm{\Sigma}^{-1}$, and $\bm{\tilde{\beta}} = z^*\bm{\Psi}^{-1}\beta/\varrho $.
After some algebraic operations, $p_z(z|\mathcal{H}_0)$ can be written as
\begin{equation}
\begin{aligned}
&	p_z(z|\mathcal{H}_0) = \frac{\det \bm{\Psi}^{-1}}{\pi \varrho \det \bm{\Sigma}} \exp \left( -\frac{|z|^2}{\varrho} + \bm{ \tilde{\beta}}^H \bm{\Psi} \bm{ \tilde{\beta}} \right) \\
	&~~ \times  \underbrace{\int \frac{1}{\pi^K \det \bm{\Psi}^{-1}} \exp \left( -(\bm{\xi} - \bm{\tilde{\beta}})^H \bm{\Psi} (\bm{\xi} - \bm{\tilde{\beta}})   \right) {\rm d } \bm{\xi}}_{=1}.
\end{aligned}
\end{equation}
Note that the  integrand in $p_z(z|\mathcal{H}_0)$ is a Gaussian pdf, and thus  this integration equals to 1.

Next, according to the Sherman-Morrison matrix inversion lemma \cite{Add4}, we obtain
\begin{equation}
	\bm{\Psi}^{-1} = \bm{\Sigma}-\frac{\bm{\Sigma} \bm{\beta \beta}^H  \bm{\Sigma}}{\varrho + \bm{\beta}^H \bm{\Sigma} \bm{\beta} }.
\end{equation}
Therefore, we have
\begin{equation}
	\bm{ \tilde{\beta}}^H \bm{\Psi} \bm{ \tilde{\beta}} = \frac{|z|^2}{\varrho^2} \bm{\beta}^H \bm{\Psi}^{-1} \bm{\beta}=\frac{|z|^2}{\varrho^2}\frac{\bm{\beta}^H \bm{\Sigma} \bm{\beta}}{\varrho +\bm{\beta}^H \bm{\Sigma} \bm{\beta} }.
\end{equation}
Then, the exponential term in $p_z(z|\mathcal{H}_0)$ is
\begin{equation}
	-\frac{|z|^2}{\varrho} + \bm{ \tilde{\beta}}^H \bm{\Psi} \bm{ \tilde{\beta}}= - \frac{|z|^2}{\varrho +\bm{\beta}^H \bm{\Sigma} \bm{\beta} }.
\end{equation}
In addition, the determinant of $\bm{\Psi}^{-1}$ can be denoted as
\begin{equation}
\begin{aligned}
	\det \bm{\Psi}^{-1} &= \det   \bm{\Sigma} \det \left( \mathbf{I}_K - \frac{\bm{\beta \beta}^H \bm{\Sigma}}{\varrho + \bm{\beta}^H \bm{\Sigma} \bm{\beta}}    \right) \\
	&= \left( 1 - \frac{\bm{ \beta}^H \bm{\Sigma} \bm{\beta}}{\varrho + \bm{\beta}^H \bm{\Sigma} \bm{\beta}}    \right) \det \bm{\Sigma}=\frac{\varrho  \det   \bm{\Sigma}}{\varrho + \bm{\beta}^H \bm{\Sigma} \bm{\beta}}.
\end{aligned}
\end{equation}
Hence, $\varrho \det \bm{\Sigma}/ \det \bm{\Psi}^{-1} = \varrho + \bm{\beta}^H \bm{\Sigma} \bm{\beta}$.

Finally, summarizing the above derivations, produces
\begin{equation}
p_z( z| \mathcal{H}_0 ) = \frac{1}{\pi \sigma_{in }^2 } {\rm exp} \left(  - \frac{|z|^2}{ \sigma_{in }^2   }   \right),
\end{equation}
where $\sigma_{in }^2 = 2\mathbf{w}^H\mathbf{w} + \bm{\beta}^H \bm{\Sigma} \bm{\beta}=  \varrho + \bm{\beta}^H \bm{\Sigma} \bm{\beta}$. The proof is completed.

\color{black}

\section{Computations of $\mathbf{C}(\omega_k)$ and $\mathbf{D}(\omega_k)$}
Consider the normalized angle $\omega = \sin \theta$ and let $\mathbf{\tilde{a}}_t(\omega) = \mathbf{a}_t(\theta)$ and $\mathbf{\tilde{a}}_r(\omega) = \mathbf{a}_r(\theta)$. Using some matrix operations gives rise to
\begin{equation}
\begin{aligned}
	\mathbf{A}(\theta)\mathbf{s} &= \text{vec}\{\mathbf{a}_r(\theta)\mathbf{a}_t^T(\theta)\mathbf{S}  \} =  \text{vec}\{\mathbf{\tilde{a}}_r(\omega)\mathbf{\tilde{a}}_t^T(\omega)\mathbf{S}  \} \\
	&=  ( \mathbf{S}^T \otimes \mathbf{I}_{N_r}  ) \text{vec}  \{ \mathbf{\tilde{a}}_r(\omega)\mathbf{\tilde{a}}_t^T(\omega)  \} \\
	& = ( \mathbf{S}^T \otimes \mathbf{I}_{N_r}  ) (\mathbf{\tilde{a}}_t(\omega) \otimes \mathbf{\tilde{a}}_r(\omega)  ).
\end{aligned}
\end{equation}
Then
\begin{equation}
\begin{aligned}
	\mathbb{E}\{ \mathbf{A}(\theta)\mathbf{s} \mathbf{s}^H  \mathbf{A}^H(\theta) \} =  ( \mathbf{S}^T \otimes \mathbf{I}_{N_r}  ) \mathbf{C}(\omega)  ( \mathbf{S}^T \otimes \mathbf{I}_{N_r}  )^H,
\end{aligned}
\end{equation}
where 
\begin{equation}
\begin{aligned}
\mathbf{C}(\omega) &=	\mathbb{E}\{ (\mathbf{\tilde{a}}_t(\omega) \otimes \mathbf{\tilde{a}}_r(\omega)  ) (\mathbf{\tilde{a}}_t^H(\omega) \otimes \mathbf{\tilde{a}}_r^H(\omega)  ) \} \\
&=\mathbb{E}\{ \mathbf{\tilde{a}}_t(\omega)\mathbf{\tilde{a}}_t^H(\omega) \otimes \mathbf{\tilde{a}}_r(\omega)\mathbf{\tilde{a}}_r^H(\omega)   \}.
\end{aligned}
\end{equation}
Note that $\mathbf{C}(\omega) \in \mathbb{C}^{N_tN_r \times N_tN_r}$ is made up of $N_t \times N_t$ blocks, each of which is an $N_r \times N_r$ matrix.

Assume that $\omega$ obeys uniform distribution $\omega \sim \mathcal{U}(\varpi-\delta,\varpi+\delta)$.
For the $(m,n)$-th ($m,n=1,2,...,N_t$) block of  $\mathbf{C}(\omega)$, the $(p,q)$-th ($p,q=1,2,...,N_r$) element is given by
\color{black}
\begin{equation}
\label{eq67}
	\begin{aligned}
	\mathbf{C}_{(m,n)}^{(p,q)}(\omega) &= \frac{1}{N_tN_r} \int_{\varpi-\delta}^{\varpi+\delta} \frac{1}{2\delta} e^{-\jmath\pi \tilde{d}_{(m,n)}^{(p,q)} \omega } \text{d} \omega \\
	&= \frac{e^{-\jmath\pi \tilde{d}_{(m,n)}^{(p,q)} \varpi }}{N_tN_r} \text{sinc} ( \tilde{d}_{(m,n)}^{(p,q)} \delta ),
	\end{aligned}
\end{equation}
where $\tilde{d}_{(m,n)}^{(p,q)} = 2(d_{t,m}-d_{t,n} + d_{r,p} -d_{r,q})/\lambda $.
\color{black}
Considering $\omega = \omega_k$, $\omega_k \sim \mathcal{U}( \varpi_k  -\delta_k, \varpi_k  +\delta_k ),~k=1,2,...,K$ and utilizing \eqref{eq67}, we obtain $\mathbf{C}(\omega_k)$.

Similarly, to compute $\mathbf{D}(\omega_k)$, let us consider
\begin{equation}
\begin{aligned}
\mathbf{A}^H(\theta)\mathbf{w} &= \text{vec}\{\mathbf{a}_t^*(\theta)\mathbf{a}_r^H(\theta)\mathbf{W}  \} =  \text{vec}\{\mathbf{\tilde{a}}_t^*(\omega)\mathbf{\tilde{a}}_r^H(\omega)\mathbf{W}  \} \\
&=  ( \mathbf{W}^T \otimes \mathbf{I}_{N_t}  ) \text{vec}  \{ \mathbf{\tilde{a}}_t^*(\omega)\mathbf{\tilde{a}}_r^H(\omega) \} \\
& = ( \mathbf{W}^T \otimes \mathbf{I}_{N_t}  ) (\mathbf{\tilde{a}}_r^*(\omega) \otimes \mathbf{\tilde{a}}_t^*(\omega)  ).
\end{aligned}
\end{equation}
Then, 
\begin{equation}
\begin{aligned}
\mathbb{E}\{ \mathbf{A}^H(\theta)\mathbf{w} \mathbf{w}^H  \mathbf{A}(\theta) \} =  ( \mathbf{W}^T \otimes \mathbf{I}_{N_t}  ) \mathbf{D}(\omega)  ( \mathbf{W}^T \otimes \mathbf{I}_{N_t}  )^H,
\end{aligned}
\end{equation}
where 
\begin{equation}
\begin{aligned}
\mathbf{D}(\omega) &=	\mathbb{E}\{ (\mathbf{\tilde{a}}_r^*(\omega) \otimes \mathbf{\tilde{a}}_t^*(\omega)  ) (\mathbf{\tilde{a}}_r^T(\omega) \otimes \mathbf{\tilde{a}}_t^T(\omega)  ) \} \\
&=\mathbb{E}\{ \mathbf{\tilde{a}}_r^*(\omega)\mathbf{\tilde{a}}_r^T(\omega) \otimes \mathbf{\tilde{a}}_t^*(\omega)\mathbf{\tilde{a}}_t^T(\omega)   \},
\end{aligned}
\end{equation}
is composed of  $N_r \times N_r$ blocks, each of which is an $N_t \times N_t$  matrix.

For the $(m,n)$-th ($m,n=1,2,...,N_r$) block of  $\mathbf{D}(\omega)$, the $(p,q)$-th ($p,q=1,2,...,N_t$) element is given by
\color{black}
\begin{equation}
\label{eq67n}
\begin{aligned}
\mathbf{D}_{(m,n)}^{(p,q)}(\omega) &= \frac{1}{N_tN_r} \int_{\varpi-\delta}^{\varpi+\delta} \frac{1}{2\delta} e^{\jmath\pi \tilde{d}^{(m,n)}_{(p,q)} \omega } \text{d} \omega \\
&= \frac{e^{\jmath\pi \tilde{d}^{(m,n)}_{(p,q)} \varpi }}{N_tN_r} \text{sinc} ( \tilde{d}^{(m,n)}_{(p,q)} \delta ),
\end{aligned}
\end{equation}
where $\tilde{d}^{(m,n)}_{(p,q)}  = 2(d_{r,m} - d_{r,n} + d_{t,p} - d_{t,q})/\lambda$. \color{black}
Considering $\omega = \omega_k$, $\omega_k \sim \mathcal{U}( \varpi_k  -\delta_k, \varpi_k  +\delta_k )$ and utilizing \eqref{eq67n}, we obtain $\mathbf{D}(\omega_k)$.

\section{Proof of Proposition 2}
The KKT condition of the problem \eqref{subproblem_sr} can be given by
\begin{subequations}
	\label{eq68}
	\begin{align}
		&\frac{2\mathbf{\tilde{\Phi}}(\mathbf{w})\mathbf{\tilde{s}} }{ \mathbf{\tilde{s}}^T \mathbf{\tilde{\Gamma}}(\mathbf{w}) \mathbf{\tilde{s}} } - \frac{2 \mathbf{\tilde{s}}^T \mathbf{\tilde{\Phi}}(\mathbf{w}) \mathbf{\tilde{s}} \mathbf{\tilde{\Gamma}}(\mathbf{w}) \mathbf{\tilde{s}} }{  (\mathbf{\tilde{s}}^T \mathbf{\tilde{\Gamma}}(\mathbf{w}) \mathbf{\tilde{s}})^2  }  + \bm{\vartheta}_u - \bm{\vartheta}_l + 2 \nu \mathbf{\tilde{s}} = \mathbf{0}_{2N_tL}, \\
	&\mathbf{\tilde{s}}^T\mathbf{\tilde{s}}=1,~\frac{-1}{\sqrt{2N_tL}} \mathbf{1}_{2N_tL} \leq \mathbf{\tilde{s}} \leq \frac{1}{\sqrt{2N_tL}} \mathbf{1}_{2N_tL}, \\
	& \bm{\vartheta}_u \odot (\mathbf{\tilde{s}} -  \frac{1}{\sqrt{2N_tL}} \mathbf{1}_{2N_tL} ) =\mathbf{0}_{2N_tL},~\bm{\vartheta}_u\geq \mathbf{0}_{2N_tL}, \label{68c} \\
	&  \bm{\vartheta}_l \odot (\mathbf{\tilde{s}} +  \frac{1}{\sqrt{2N_tL}} \mathbf{1}_{2N_tL} ) =\mathbf{0}_{2N_tL},~\bm{\vartheta}_l\geq \mathbf{0}_{2N_tL}, \label{68d} 
	\end{align}
\end{subequations}
where $\nu$, $\bm{\vartheta}_u$, and $\bm{\vartheta}_l$ are dual variables.

According to the update rule in \eqref{eq26}, firstly, $\mathbf{\tilde{s}}^{(j+1)}$ is obtained by minimizing $\frac{\rho_1}{2}||\mathbf{t}^{(j)}-\mathbf{\tilde{s}}+\mathbf{u}_1^{(j)}||^2_2+\frac{\rho_2}{2}||\mathbf{r}^{(j)}-\mathbf{\tilde{s}}+\mathbf{u}_2^{(j)}||^2_2$ subject to $\mathbf{\tilde{s}} \in \mathbb{D}$, and thus  $\mathbf{\tilde{s}}^{(j+1)}$  satisfies
	\begin{equation}
	\label{eq69}
	\begin{aligned}
			\rho_1 (\mathbf{\tilde{s}}^{(j+1)} - \mathbf{t}^{(j)}-\mathbf{u}_1^{(j)}   )
	& + \rho_2 (\mathbf{\tilde{s}}^{(j+1)} - \mathbf{r}^{(j)}-\mathbf{u}_2^{(j)}   ) \\
		&	+ \bm{\vartheta}_u^{(j+1)}- \bm{\vartheta}_l^{(j+1)} =\mathbf{0}_{2N_tL},
	\end{aligned}
	\end{equation}
where $\bm{\vartheta}_u^{(j+1)}$ and $\bm{\vartheta}_l^{(j+1)}$ satisfy \eqref{68c} and \eqref{68d}.

Secondly, subject to $\mathbf{t}^T\mathbf{t}=1$, $\mathbf{t}^{(j+1)}$ is updated by solving $\min\limits_{\mathbf{t}}~\mathcal{L}_{\rho_1,\rho_2}(\mathbf{\tilde{s}}^{(j+1)},\mathbf{t},\mathbf{r}^{(j)},\mathbf{u}_1^{(j)},\mathbf{u}_2^{(j)})$.
Therefore, we have
\begin{equation}
	\label{eq70}
	\begin{aligned}
	\frac{2\mathbf{\tilde{\Phi}}(\mathbf{w})\mathbf{t}^{(j+1)} }{ (\mathbf{r}^{(j)})^T \mathbf{\tilde{\Gamma}}(\mathbf{w}) \mathbf{r}^{(j)} } &+\rho_1(\mathbf{t}^{(j+1)}-\mathbf{\tilde{s}}^{(j+1)} +\mathbf{u}_1^{(j)} ) \\
	&+ 2\nu^{(j+1)} \ \mathbf{t}^{(j+1)} =\mathbf{0}_{2N_tL}.
	\end{aligned}
\end{equation}

In addition, $\mathbf{r}^{(j+1)}$ is obtained by solving the unconstrained problem $\min\limits_{\mathbf{r}}~\mathcal{L}_{\rho_1,\rho_2}(\mathbf{\tilde{s}}^{(j+1)},\mathbf{t}^{(j+1)},\mathbf{r},\mathbf{u}_1^{(j)},\mathbf{u}_2^{(j)})$, which yields
\begin{equation}
	\label{eq71}
\begin{aligned}
 &- \frac{2 (\mathbf{t}^{(j+1)})^T \mathbf{\tilde{\Phi}}(\mathbf{w}) \mathbf{t}^{(j+1)} \mathbf{\tilde{\Gamma}}(\mathbf{w}) \mathbf{r}^{(j+1)} }{  ((\mathbf{r}^{(j+1)})^T \mathbf{\tilde{\Gamma}}(\mathbf{w}) \mathbf{r}^{(j+1)})^2  } \\
 &~~~~~~~~~+ \rho_2(\mathbf{r}^{(j+1)}-\mathbf{\tilde{s}}^{(j+1)} + \mathbf{u}_2^{(j)}  )=\mathbf{0}_{2N_tL}.
\end{aligned}
\end{equation}

Finally, combining the equations \eqref{eq69}-\eqref{eq71} and using the condition $\mathbf{d}^{(j+1)},\mathbf{c}_1^{(j+1)},\mathbf{c}_2^{(j+1)} \rightarrow \mathbf{0}_{2N_tL}$, we have 
\begin{equation}
	\begin{aligned}
	&\frac{2\mathbf{\tilde{\Phi}}(\mathbf{w})\mathbf{\tilde{s}}^{(j+1)} }{ (\mathbf{\tilde{s}}^{(j+1)})^T \mathbf{\tilde{\Gamma}}(\mathbf{w}) \mathbf{\tilde{s}}^{(j+1)} } -\frac{2 (\mathbf{\tilde{s}}^{(j+1)})^T \mathbf{\tilde{\Phi}}(\mathbf{w}) \mathbf{\tilde{s}}^{(j+1)} \mathbf{\tilde{\Gamma}}(\mathbf{w}) \mathbf{\tilde{s}}^{(j+1)} }{  ((\mathbf{\tilde{s}}^{(j+1)})^T \mathbf{\tilde{\Gamma}}(\mathbf{w}) \mathbf{\tilde{s}}^{(j+1)})^2  } \\
	&~~~~~~~~~~~~+\bm{\vartheta}_u^{(j+1)}- \bm{\vartheta}_l^{(j+1)} + 2\nu^{(j+1)} \ \mathbf{\tilde{s}}^{(j+1)} =\mathbf{0}_{2N_tL}.
	\end{aligned}
\end{equation}
As a result, $\mathbf{\tilde{s}}^{(j+1)}$ satisfies the KKT condition \eqref{eq68}, and the proof is completed.



%




%

\balance
\bibliographystyle{IEEEtran}
\bibliography{IEEEabrv,A}

%

%
%
%




\end{document}